\documentclass[useAMS, usenatbib]{mn2e}
\usepackage{aas_macros}
\usepackage{amsmath}
\usepackage{amssymb} 
\usepackage{graphicx} 
\usepackage{float} 
\usepackage{amsfonts}
\usepackage{lmodern}
\usepackage{bm}
\usepackage{placeins}

\newcommand{\bc}{\begin{center}}
\newcommand{\ec}{\end{center}}

\let\oldhat\hat
\renewcommand{\hat}[1]{\oldhat{\mathbf{#1}}}
\let\oldbullet\bullet \renewcommand{\bullet}[1][0pt]{% 
\mathrel{\raisebox{#1}{$\oldbullet$}}%
}
\title[Cosmological simulations of dwarfs]{Cosmological simulations of dwarfs: the need for ISM physics beyond SN
feedback alone} \author[Smith, Sijacki \& Shen]{Matthew C. Smith$^{1,2}$\thanks{E-mail: m.c.smith@ast.cam.ac.uk}, Debora
  Sijacki$^1$, Sijing Shen$^{3}$ \\
  $^1$ Institute of Astronomy and Kavli Institute for Cosmology,
  University of Cambridge, Madingley Road, Cambridge CB3 0HA, UK \\
  $^2$ Center for Computational Astrophysics, Flatiron Institute, 162 5\textsuperscript{th} Avenue, New York, NY 10010, USA \\
  $^3$ Institute of Theoretical Astrophysics, University of Oslo, P.O. Box 1029, Blindern, N-0315, Oslo, Norway}

\begin{document}

\maketitle

\begin{abstract}
The dominant feedback mechanism in low mass haloes is usually assumed to take
the form of massive stars exploding as supernovae (SNe). We perform very high
resolution cosmological zoom-in simulations of five dwarf galaxies to $z = 4$
with our mechanical SN feedback model. This delivers the correct amount of
momentum corresponding to the stage of the SN remnant evolution resolved,
and has been shown to lead to realistic dwarf properties in isolated simulations. We find that
in 4 out of our 5 simulated cosmological dwarfs, SN feedback has insufficient
impact resulting in excessive stellar masses, extremely compact sizes and
central super-solar stellar metallicities. The failure of the SN feedback in
our dwarfs is physical in nature within our model and is the result of the build up of very
dense gas in the early universe due to mergers and cosmic inflows prior to the
first SN occurring. We demonstrate that our results are insensitive to resolution (provided that it is high enough), 
details of the (spatially uniform) UV background and reasonable alterations within our star formation prescription. 
We therefore conclude that the ability of SNe to regulate dwarf galaxy properties is dependent on other physical 
processes, such as turbulent pressure support, clustering and runaway of SN progenitors and other sources of 
stellar feedback. 
\end{abstract}


\section{Introduction} \label{intro}
While they may be the least massive and luminous systems in our universe,
understanding the origin of dwarf galaxy properties represents a key step
in developing and testing theories of galaxy formation and cosmology.
From the perspective of cosmology, abundances and structural properties of low mass
haloes present an important observational test of the $\Lambda$CDM model,
but the effects of baryonic physics (to which dwarfs are highly susceptible because
of their small potential wells) can make it difficult to make robust
predictions. Meanwhile, from the point of view of galaxy formation, these
`messy' baryonic processes are interesting in their own right, as well as providing
insight into the reionization history of the universe and the enrichment of the intergalactic
medium (IGM) with metals.

Examining the issue of dwarf abundances, there is a substantial offset between the predicted
dark matter halo abundance from numerical simulations and the observed galaxy stellar mass
function \citep[for recent work see e.g.][]{Behroozi2013,Moster2018},
indicating that dwarfs must be more than 
an order of magnitude less efficient at forming stars than Milky Way-sized
haloes. This is also posited as a solution to the so called `missing satellites
problem', where the observed number of Local Group satellites is at odds with the substantially
larger number of dark matter haloes predicted by cosmological N-body
simulations \citep[see e.g.][]{Moore1999, Klypin1999, Diemand2008,
  Springel2008, Koposov2009, Rashkov2012, Sawala2016}.  

It has been suggested for some time that low mass
haloes should have their star formation efficiency suppressed by two primary
processes: SN feedback 
\citep[e.g.][]{Larson1974, Dekel1986, Mori2002, Governato2007} and cosmic
reionization \citep[e.g.][]{Efstathiou1992, Bullock2000, Dijkstra2004,
  Kravtsov2004, Madau2008}. Evidence that these structures do in fact exist
but are relatively dark has been bolstered recently by detections of local
ultra-faint dwarfs \citep[e.g.][]{Koposov2015, Laevens2015, Martin2015, KimD2015}.

As well as influencing abundances of low mass haloes, baryonic physics has also been invoked to
solve structural discrepancies between dark matter simulations and
observations. One such discrepancy is often termed the `cusp-core
controversy'. Within the $\Lambda$CDM paradigm, dark matter-only simulations systematically 
predict steep inner density profiles for these low mass haloes, but some
observations suggest that they may instead contain low density cores \citep[see
  e.g.][]{Moore1994, Flores1994, deBlok2002, Walker2011}. While SN feedback
has been widely invoked in hydrodynamical simulations in an attempt to generate
cored density profiles, there is still no consensus in the literature as some
groups find only cuspy profiles 
 \citep[e.g.][]{Vogelsberger2014, Sawala2016}, while others find various
 levels of cored profiles with different trends as a function of halo mass or
 redshift \citep[e.g.][]{Navarro1996a, Gnedin2002, Read2005, Mashchenko2008,
   Governato2010, Pontzen2012, DiCintio2014, Onorbe2015, Fitts2017}. The level
 of success in transforming dark matter cusps into cores seems closely
 related to the degree of burstiness of SN feedback, which could also affect
 the mass-loading of galactic outflows and the early enrichment of the IGM.

 It is perhaps at some level unsurprising that the properties of simulated dwarfs
 predicted by different groups are at variance as very different sub-grid
 models for star formation, SN feedback and wind launching are adopted, in
 addition to results often being rather sensitive to the numerical resolution
 of the simulations. This is however clearly unsatisfactory if we are to
 understand at a more fundamental level how SN feedback operates in dwarf
 galaxies, and even more so if we are to derive robust constraints on the
 nature of dark matter, using observed dwarfs as near-field cosmology probes.
   
Recently, based on analytical calculations or small scale simulations of individual SN
explosions, there have been several theoretical works (e.g. \citealt{Hopkins2014a,
  Kimm2014, Iffrig2015, Martizzi2015, Kim2015, Walch2015}, but see
also earlier work by \citealt{Cioffi1988,Thornton1998}) 
aiming at quantifying the
correct momentum injection at a given SN remnant stage, as a function of local
ISM properties, such as the gas density, metallicity and porosity. These
studies are particularly useful as they in principle allow the imparting of the
appropriate momentum into the ISM, even when the Sedov-Taylor phase of the SN
remnant evolution is not properly resolved (often the case in galaxy formation
simulations), without the use of tunable parameters (although they usually make 
certain assumptions such as a uniform ambient medium).
We have trialled this type of SN injection, often dubbed `mechanical feedback',
in an extensive series of simulations of isolated dwarf galaxies
\citep{Smith2018}, finding that it results in realistic and well converged star formation
rates and morphologies over two orders of magnitude in mass resolution. 
However, to naturally produce mass-loaded, multi-phase outflows gas mass
resolution needed to be of order of few tens of $\mathrm{M_\odot}$ at
least. 

Thankfully, even though these resolution requirements are quite
daunting, they are achievable in full cosmological simulations provided that
one dwarf is simulated at a time with a zoom-in technique. Hence, the aim of
this work is to examine the mechanical SN feedback scheme in
fully self-consistently formed dwarfs without tuning any parameters to
understand if it leads to realistic dwarf properties once the cosmological gas
inflows and mergers are taken into account. We explore this by
randomly selecting five dwarfs with virial masses between $\sim 2 -6 \times 10^9
\,\mathrm{M_\odot}$ at $z=4$ which reside in different environments and have 
different assembly histories. 

\section{Methodology}\label{Methodology}
Our numerical scheme is essentially the same as that described in
\cite{Smith2018}, but we summarise the salient details here. We carry out our
simulations with the moving-mesh code \textsc{Arepo} \citep{Springel2010}
which solves hydrodynamics on an unstructured Voronoi mesh. Gravity is
included using a hybrid TreePM scheme. 

In this work, we include radiative cooling as in \cite{Vogelsberger2013}.
Primordial heating and cooling rates are calculated using cooling, recombination
and collisional rates provided by \cite{Cen1992} and \cite{Katz1996}. Metal-line
cooling to 10~K is obtained from lookup tables containing rates precalculated from
the photoionization code \textsc{Cloudy}. We include a redshift dependent,
but spatially homogeneous UV background from \cite{FG2009}, although it is only
turned on from $z=9$ to approximate the latest Planck measurement of optical
depth to reionization \citep{Planck2016}\footnote{However, we have performed extra
simulations where the UV background follows \cite{FG2009} exactly (switching on
at $z=11.7$) and find that it does not change our results in any appreciable way.}. 
We adopt the density based
self-shielding prescription of \cite{Rahmati2013} to attenuate the UV background
in dense gas.

\begin{figure*}
\centering
\includegraphics[trim=0 7 0 0,clip]{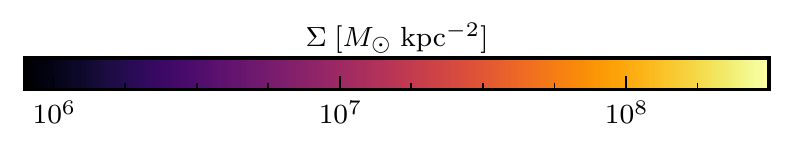}
\includegraphics[trim=0 0 0 5,clip,width=1.0\textwidth]{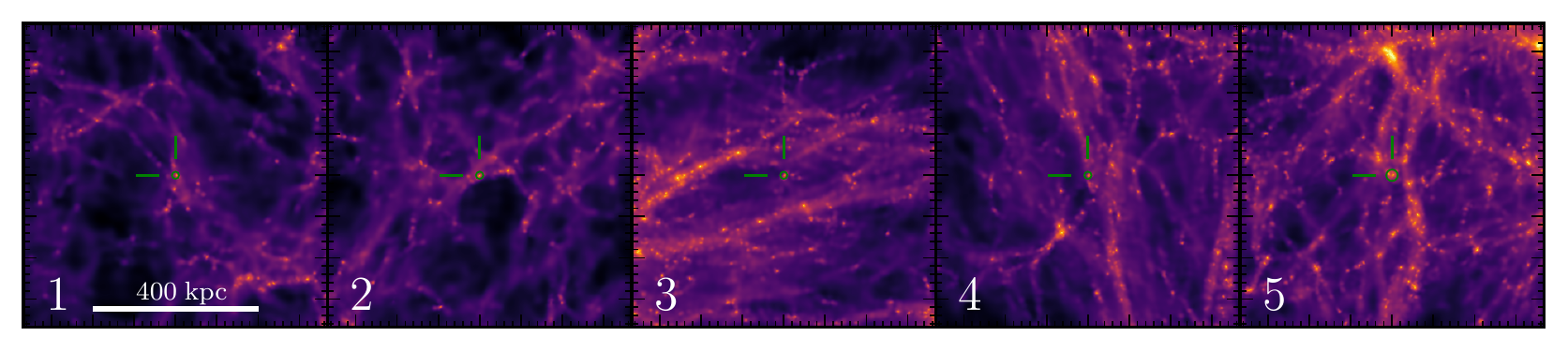}
\caption{Density projections of the large scale environment around our target
  haloes in the coarse dark matter only simulation at $z = 4$. The target
  haloes are marked with green ticks for ease of identification and the virial
  radius is marked with a green circle. We deliberately select the haloes
    from a variety of environments ranging from void-like regions to rich
    filaments.}  
\label{projections_wide} 
\end{figure*}

We include a non-thermal pressure floor to prevent artificial fragmentation in
the event of under-resolving the Jeans length \citep[see e.g.][]{Truelove1997}.
To ensure that the Jeans length is resolved by $N_\mathrm{J}$ cells, this takes
the form

\begin{equation}
P_\mathrm{min} = \frac{N_\mathrm{J}^2 \Delta x^2 G \rho^2}{\pi \gamma}\,, 
\label{pressure_floor}
\end{equation}
where $\Delta x$ is the cell diameter, $\rho$ is the gas density and 
$\gamma = 5/3$ is the adiabatic index. We adopt $N_\mathrm{J}=8$ in this work.
A detailed discussion of the effects of adopting a pressure floor can be found
in \cite{Smith2018} (see also Section~\ref{Section_parameters}).

Gas above a density threshold of $n_\mathrm{SF}$ is assigned a star formation
rate density according to a simple Schmidt law,

\begin{equation}
\dot{\rho}_{*} = \epsilon_\mathrm{SF}\frac{\rho}{t_{\mathrm{ff}}}\,, 
\label{eq:schmidt}
\end{equation}
where $\rho$ is the gas density, $\epsilon_\mathrm{SF}$ is some efficiency and 
$t_{\mathrm{ff}}=\sqrt{3\pi/32G\rho}$ is the free-fall time. We use a fiducial 
value of $n_{\mathrm{SF}}=10\ \mathrm{cm}^{-3}$ and $\epsilon_\mathrm{SF}=1.5\%$ 
\citep[chosen to match observed efficiencies in dense gas, see e.g.][and
  references therein]{Krumholz2007}. 
We also examine the effect of varying these parameters in Section~\ref{Section_parameters}.
Using these rates, gas cells are then stochastically converted into star particles 
(collisionless particles representing single stellar populations). Star particles
inherit the metallicity of the gas from which they were formed.

For each star particle, we obtain a SN rate, $\dot{N}_\mathrm{SN}$, as a function of
age and metallicity precalculated using \mbox{\textsc{Starburst99}} \citep{Leitherer1999} assuming
a \cite{Kroupa2002} IMF. The number of SNe that occur in a timestep is then drawn
from a Poisson distribution with a mean of $\bar{N}_\mathrm{SN}=\dot{N}_\mathrm{SN}\Delta t$,
where $\Delta t$ is the timestep. In order to individually time resolve SNe, we impose
a timestep limiter for star particles to ensure that $\bar{N}_\mathrm{SN}\ll1$.

When a SN occurs, mass, metals, energy and momentum are coupled to the gas cell containing
the star particle (the host cell) and its neighbours (all those that share a face with the host cell).
Feedback quantities are distributed to the gas cells using an explicitly isotropic weighting scheme
in the rest frame of the star particle (details in \citealt{Smith2018}, see also \citealt{Hopkins2018})
in order to avoid spurious numerical effects that can arise when using a simple kernel (mass) weighted
approach to nearest neighbours due to the increased relative number of resolution elements present in
dense gas. The ejecta mass, $m_\mathrm{ej}$, deposited per SN is $10\ \mathrm{M_\odot}$, of which 
$2\ \mathrm{M_\odot}$ is in metals, with an energy of $10^{51}\ \mathrm{ergs}$. In simulations 
designated `no feedback', mass and metals are returned, but no feedback energy/momentum is deposited. 
In runs with full SN feedback, we adopt the mechanical feedback scheme described in \cite{Smith2018} 
\cite[see also][]{Hopkins2014a,Hopkins2017a,Hopkins2018,Kimm2014,Kimm2015,Martizzi2015}. This aims to 
deposit the correct amount of momentum corresponding to the stage of the SN remnant evolution resolved
(dependent on the local gas density and metallicity).

When analysing simulations, we use the halo finder \textsc{Subfind}
\citep{Springel2001,Dolag2009} to calculate halo properties.  We adopt the convention of considering friends-of-friends (FOF) groups as the primary dark matter
halo and subhaloes as galaxies within the halo (unless otherwise stated, we only consider centrals). For halo
virial masses, we use the definition of \cite{Bryan1998} and for galaxy stellar masses we use the mass
contained within twice the radius that contains half the total subhalo stellar mass associated with the group.
We use the code \textsc{Sublink} \citep{Rodriguez-Gomez2015} to construct merger trees and track haloes/subhaloes
throughout the simulations. We follow the branch of the merger tree with the most mass behind it for our analysis except where otherwise mentioned. We adopt a \cite{Planck2016} cosmology throughout this work. Unless otherwise stated, all units are in proper coordinates.
\begin{figure*}
\centering
\includegraphics[trim=0 6 0 0,clip]{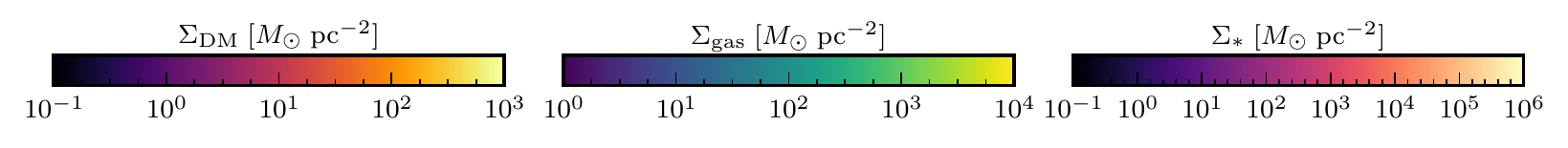}
\includegraphics[trim=0 0 0 6,clip, width=1.0\textwidth]{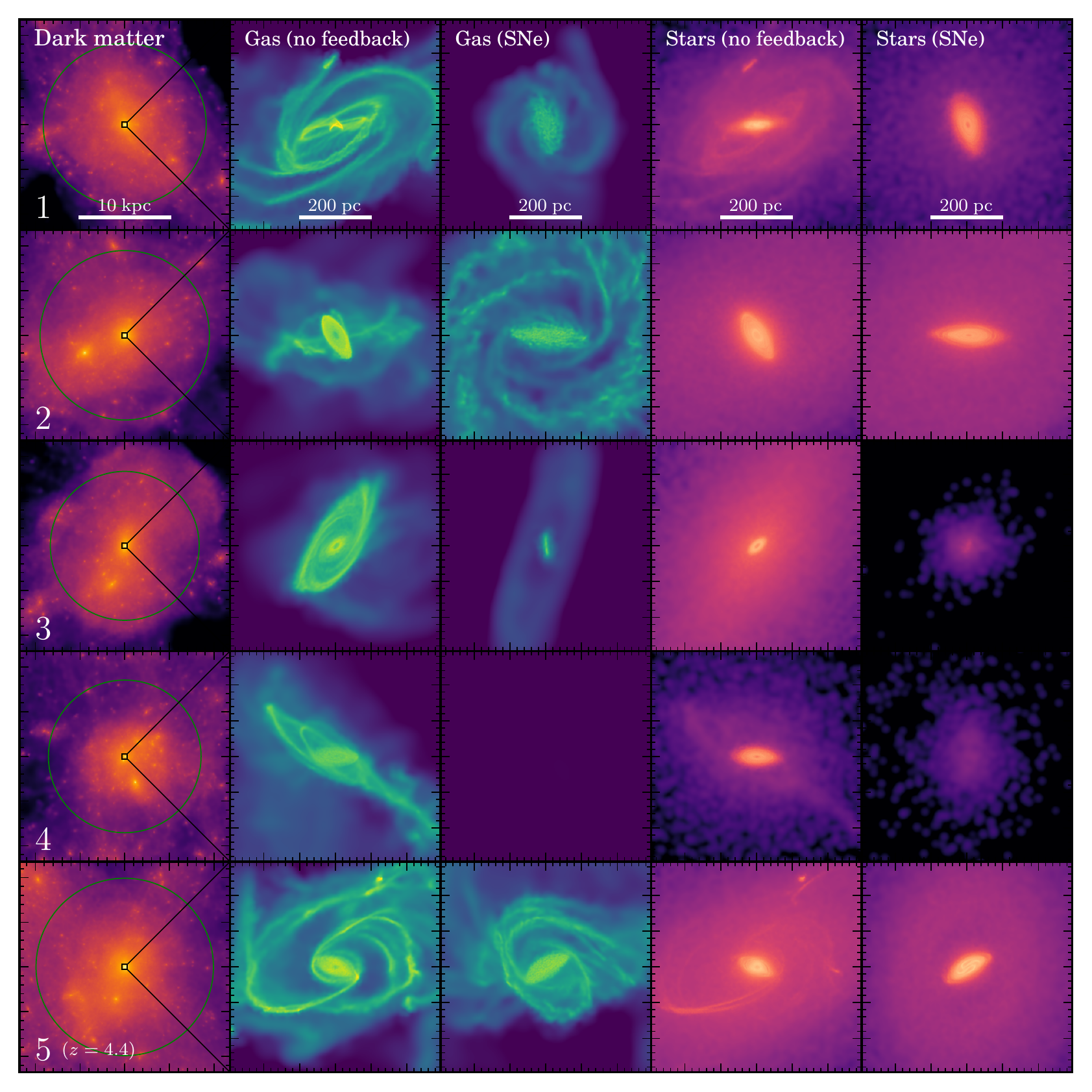}
\caption{Density projections, from left to right: dark matter (shown here for the no feedback simulations, although the equivalent plots for runs with SNe are similar), gas for simulations without SNe, gas with SNe, stars without SNe and stars with SNe. Each row corresponds to a different dwarf. The gas and stellar projections are centred on the central galaxy of the halo. Dwarfs 1, 2, 3 and 4 are shown at $z=4$, while dwarf 5 is shown at $z=4.4$ to allow comparison to the curtailed no feedback simulation. The virial radius is indicated with a green circle. While SN feedback significantly alters morphologies, particularly in the case of dwarf 4, in most cases centrally condensed baryon concentration persists.}  
\label{projections} 
\end{figure*}
\begin{figure*}
\centering
\includegraphics[width=1.0\textwidth]{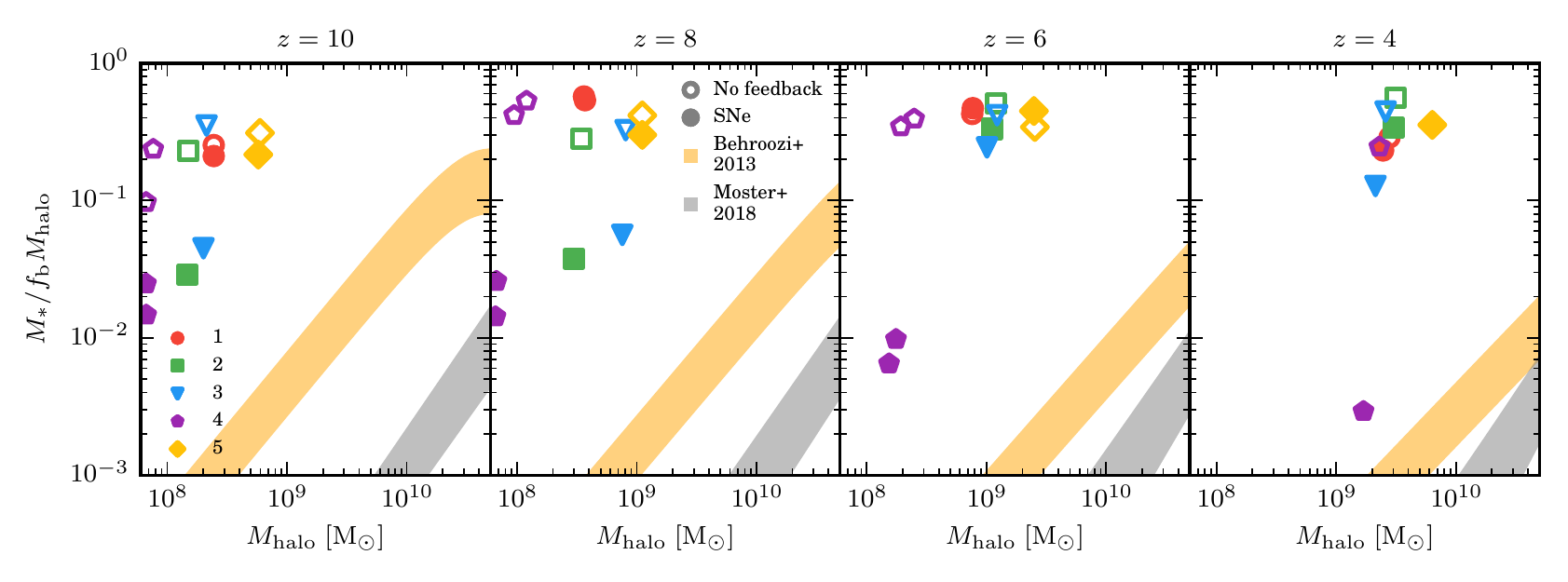}
\caption{Stellar mass to halo mass ratio as a function of halo mass for our
  various simulations at $z=10$, 8, 6 and 4. Halo mass here is defined as in
  \protect\cite{Bryan1998}. We express the stellar mass to halo mass ratio
  (which can be considered the integrated baryon conversion efficiency) as the
  mass of stars formed in the central galaxy (defined as the most massive
  subhalo) divided by the product of the halo mass and the cosmic baryon
  fraction. Open symbols indicate simulations without SNe. In the case of
  dwarf 4, we plot both the progenitor haloes of the final $z=4$ halo prior to
  their major merger at $z=5.5$. At $z=4$, while the two haloes have merged,
  the two central galaxies of the progenitors have not yet merged into a
  single subhalo (see main text); nonetheless, we use the sum of the stellar
  mass in both of these galaxies to compute the star formation efficiency. We
  also indicate results from abundance matching as in
  \protect\cite{Behroozi2013} and \protect\cite{Moster2018} with shaded
  regions, although it should be noted that at such low halo masses the
  relations are heavily extrapolated. Note that with exception of dwarf 4, all
  of our simulated dwarfs are in large disagreement with the abundance matching
  extrapolations.}   
\label{mhalo_star} 
\end{figure*} 

\section{Simulations}
\subsection{Initial conditions and simulation details}
The process for creating cosmological `zoom-in' initial conditions is as follows. First, a coarse resolution
dark matter only simulation of a large, periodic cosmological volume is run to a target redshift, $z_\mathrm{target}$. Dark matter
haloes of interest are identified in the $z_\mathrm{target}$ output of this simulation and are resimulated at a higher resolution with a
`zoom-in' technique. Gas is then added to the
initial conditions by splitting the particles into dark matter and gas mesh generating points according
to the cosmic baryon fraction (although we also carry out dark matter only zoom-in simulations).

We use the code \textsc{MUSIC} \citep{Hahn2011} to generate the initial conditions (at $z=127$) for both the coarse and
zoom-in simulations. Dwarfs 1 and 2 are selected
at $z = 0$ from $10\ \mathrm{cMpc}\,\mathrm{h}^{-1}$ coarse boxes with a resolution of $256^3$ particles (giving a particle 
mass of $7.47\times10^{6}\ \mathrm{M_\odot}$). In the coarse simulation, their virial masses at $z = 0$ are 
$1.04\times10^{10}\ \mathrm{M_\odot}$ and $1.12\times10^{10}\ \mathrm{M_\odot}$ with virial radii of 62.0~kpc and 64.5~kpc, respectively. The selection regions at $z=0$ are a sphere of radius 736~kpc for dwarf 1 and a sphere of radius 295~kpc for dwarf 2. 
However, for the purposes of this work, we carry out our analysis up until $z = 4$,
at which point their masses are $2.82\times10^{9}\ \mathrm{M_\odot}$ and $3.11\times10^{9}\ \mathrm{M_\odot}$, and they have virial radii of 9.20~kpc and 9.47~kpc (note that in the zoom-in simulations, the final mass and radius varies due to the effects of baryonic physics and the higher resolution). 

Dwarfs 3, 4 and 5 are selected at $z = 4$
from a $20\ \mathrm{cMpc}\,\mathrm{h}^{-1}$ box with a resolution of $512^3$ particles (i.e. the same mass resolution as the boxes
used for dwarfs 1 and 2). The masses of dwarfs 3 and 4 at $z=4$ are $2.56\times10^{9}\ \mathrm{M_\odot}$ and $2.51\times10^{9}\ \mathrm{M_\odot}$ with virial radii of 8.86~kpc and 8.78~kpc. The selection regions at $z=4$ are spheres of radius 44.1~kpc. In the coarse simulation, we identify a fifth halo with a virial mass of $1.00\times10^{10}\ \mathrm{M_\odot}$ and a virial radius of 13.96~kpc, which we resimulate with a zoom-in region of 88.3~kpc. In the subsequent zoom-in simulation, this region actually contains two separate haloes of $\sim6\times10^9\ \mathrm{M_\odot}$, sufficiently separated as to be considered independent. We take the larger of these two haloes to be the focus of our analysis, referring to it as dwarf 5. Fig.~\ref{projections_wide} shows dark matter density projections
of the large scale region around the target haloes in the coarse simulations at $z=4$. Dwarfs 1, 2 and 3 are in
relatively low density environments, 4 is in a more crowded filament region, while 5 is a larger system in a very
crowded filament.

Our fiducial simulations increase the number of resolution elements in the
zoom-in region by a factor of $16^3$
giving dark matter particle and target gas cell masses of $1536\ \mathrm{M_\odot}$ and $287\ \mathrm{M_\odot}$, respectively. We also run simulations
of dwarf 1 with a higher resolutions of $35.9$ and $15\ \mathrm{M_\odot}$ gas cell mass for the purposes of testing convergence. The
refinement/derefinement scheme in \textsc{Arepo} keeps gas cell masses within a factor of 2 of the target
mass. Because star particles are formed by converting gas cells, this also corresponds to the initial star
particle mass (prior to mass loss from feedback). We use comoving gravitational softenings of 
 0.129~ckpc for the high resolution dark matter particles, gas cells\footnote{For gas cells, the softening is calculated as the maximum of either this fixed softening value or 2.5 times the cell radius.} and star particles. For dwarfs 3, 4 and 5 the softening is held at its $z=6$ proper length of 18.4~pc from that redshift onwards, although this makes very little practical difference.
Because our simulations do not include the necessary physics (such as molecular cooling) to resolve Population III stars and the first enrichment of the ISM,
we impose a metallicity floor of $10^{-4}\ \mathrm{Z_\odot}$\footnote{Our choice is motivated by the critical metallicity
for fragmentation such that a Population II cluster can be formed \citep[see e.g.][]{Schneider2012}. This choice is somewhat arbitrary, but we find that increasing the floor value by an order of magnitude has negligible impact on our results. In practice, our haloes are rapidly enriched above the floor value.}. 
For each dwarf, we carry out a simulation to $z=4$ with no feedback and with SNe (due to computational expense, we run the no feedback dwarf 5 simulation to $z=4.4$). Section~\ref{zDiscussion} contains details of additional simulations of dwarf 1 carried out with various modifications to our fiducial parameters to test convergence.

\subsection{Results}
Fig.~\ref{projections} shows density projections of the dark matter, gas and stars for our simulated dwarfs at $z=4$ (dwarf 5 is shown at $z=4.4$ to allow comparison between the runs with and without feedback). A variety of morphologies are present. Runs without feedback tend to produce highly compact gas and stellar discs. Recent mergers can give rise to warped structures, for example in dwarfs 1 and 4. With feedback, dwarfs 1, 2 and 5 also feature compact stellar discs similar to the no feedback simulations, although their orientation has changed. The gas morphology is more obviously changed, with more irregular and diffuse structure. In dwarfs 3 and 4, feedback has made a significant impact on the stellar structure, with significantly lower surface densities and the absence of a well defined disc. In dwarf 3, most of the gas been cleared away, leaving a small dense core, while the feedback has almost completely evacuated the gas from dwarf 4.

\begin{figure*}
\centering
\includegraphics[width=1.0\textwidth]{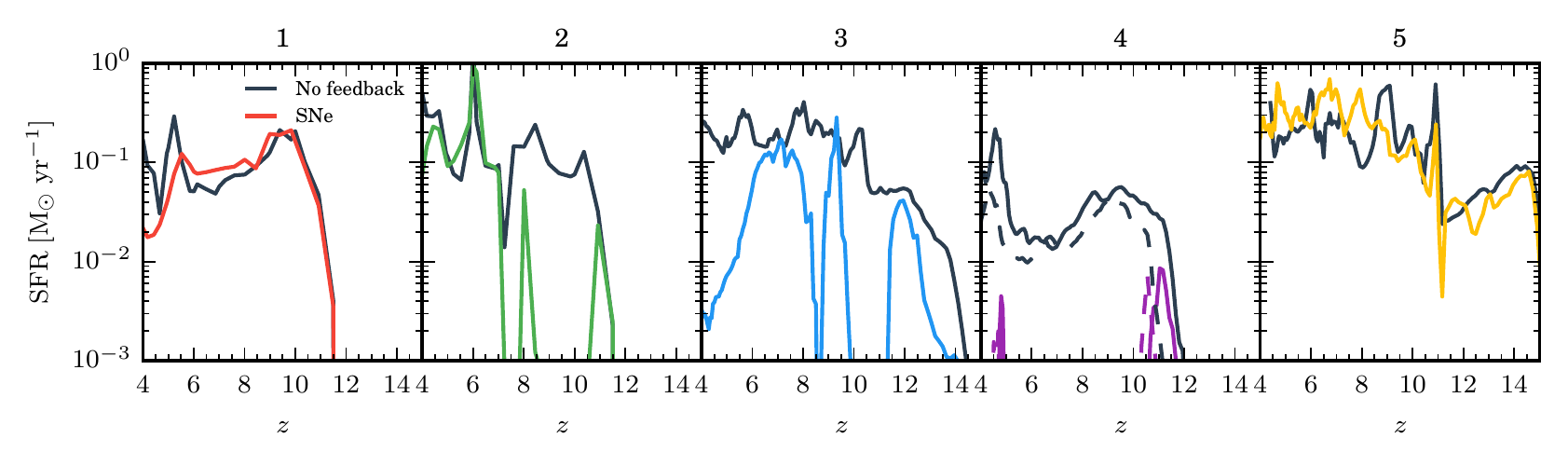}
\caption{Star formation rates as a function of redshift for the central galaxies. For dwarf 4, the central galaxies of the two almost equal mass progenitor haloes are shown, including when the two subhaloes are present in the same halo after the merger (see main text for more details); the second of the two haloes is plotted with a dashed line. Mechanical feedback in general leads to more bursty star formation rates.}
\label{sfr} 
\end{figure*}

Fig.~\ref{mhalo_star} shows the stellar mass to halo mass ratio of the central
galaxies formed in the simulations, normalised by the cosmic baryon fraction,
as a function of halo mass, for 4 redshifts ($z=10$, 8, 6, 4). We also plot
empirically derived abundance matching results from \cite{Behroozi2013} and
\cite{Moster2018} for comparison, although we have heavily extrapolated the
results to reach this mass range so they should be treated with
caution. However, even with this caveat in mind, it can be seen that the
majority of our simulated galaxies massively overproduce stars, lying several
orders of magnitude above the abundance matching relations at all four
redshifts. This is true for all simulations without SNe, where typically $10 -
60\%$ of the available baryons (taking that to be
$f_\mathrm{b}M_\mathrm{halo}$) have been converted into stars, with variation
of only a factor of a few between $z=10 - 4$. With feedback, there are mixed
results. Dwarf 1 produces almost identical stellar to halo mass ratios with
and without feedback at all redshifts, with only marginal suppression of star
formation by $z=4$. Similarly, in dwarf 5 feedback has little impact on the
evolution of stellar mass. For dwarf 2, at $z=10$, the ratio is about an order
of magnitude lower in the run with feedback than without (although still
somewhat high). However, the ratio increases with decreasing redshift and by
$z=4$ the difference is slight. Dwarf 3 has a similar behaviour to dwarf 2,
except its ratio drops relative to the no feedback simulation, eventually
lying a factor of a few lower. 

Dwarf 4 is the only case where there is a dramatic suppression of star
formation by feedback. This object has a major merger (with a ratio $\sim1.5$)
around $z=5.5$, so we treat the two progenitor haloes separately prior to
their merger. Without feedback, both progenitor haloes have similar stellar to
halo mass ratios to the other dwarfs, although they are individually of lower
mass. With the inclusion of SN feedback, the ratio is dropped by approximately
an order of magnitude at $z=10$ and this offset increases with time. By $z=4$,
the progenitor haloes have merged according to the halo finder, although the
central galaxies of the progenitors have not yet merged. For consistency, we
now calculate the stellar to halo mass ratio for the final halo by considering
the stellar mass of both of these galaxies. The ratio is now a factor of
$\sim100$ lower than the simulation without feedback and is close to the
abundance matching relations (bearing in mind their uncertainties at this
mass). The reason for the increased effectiveness of the feedback in this case
would appear to be that this object has evolved for most of its history as two
independent systems that are less massive at a given redshift than the other
simulated dwarfs, the shallower potential well increasing the relative
efficiency of the SNe to clear gas. This suggests that haloes in this mass
range are very sensitive to the manner of their assembly.

Having discussed the integrated efficiency of star formation, we now consider
the star formation histories of our dwarfs shown in Fig.~\ref{sfr}. For dwarf
4, we plot results for both central galaxies as in Fig.~\ref{mhalo_star}. For
dwarf 1, the SFRs are essentially the same in the runs with and without
SNe. Star formation starts at around $z=11.5$ and rapidly climbs to
$0.2\ \mathrm{M_\odot\ yr^{-1}}$ by $z=10$. This rapid rise in star
formation coincides with a merger at $z\sim11$. The SFRs remain
around this level until $z=4$ in the no feedback run, apart from a
merger-driven increase at $z\approx5.5$. The results of this merger are
apparent in the highly disrupted gas and disc structure visible in
Fig.~\ref{projections}. With SNe, the brief increase in SFR is arrested by the
feedback and dropped well below the no feedback rates. This burst of feedback
is responsible for the more diffuse gas apparent in
Fig.~\ref{projections}. The ability of the feedback to be effective during the
later merger but not during the first merger is due to the amount of gas
available. The subhalo gas fraction (relative to the total mass) at $z=5.5$ is
approximately a quarter of that at $z=11.5$. 

In contrast to dwarf 1, SNe are able to suppress the SFR significantly in
dwarf 2 above $z=6$. Without feedback, the SFR rises in a similar manner
to dwarf 1, although not as rapidly. However, SNe are able to restrict star
formation to a brief burst at around $z=11.5$ and another at $z=8$. It would
appear that the calmer environment (i.e. no major merger), relative to dwarf
1, at the onset of star formation allows the SNe to be effective. Dwarf 2
experiences a gas rich merger around $z=6$ that leads to a large spike in SFR
in both no feedback and feedback runs and the rapid build up of gas
overwhelming the feedback. Following this event, the SFR remains high in both
runs, leading to the similar (high) stellar mass to halo mass ratio at
$z=4$. A burst of efficient feedback around $z=4.5$ leads to a slight drop in
SFR relative to the no feedback simulations, the results of which can be seen
in the gas morphology in Fig.~\ref{projections}.

In dwarf 3, without feedback, the SFR rises slowly from $z=14$, before
becoming reasonably steady at a few $10^{-1}\ \mathrm{M_\odot\ yr^{-1}}$ from
$z=10$ onwards. This dwarf experiences no mergers of consequence, growing more
slowly than dwarfs 1 and 2, probably as a result of being in a less dense
environment. Once SNe are included, SNe are able to suppress star formation,
but only following extended bursts of high SFRs. The feedback episodes are
able to remove gas from the centre of the halo (giving rise to the morphology
seen in Fig.~\ref{projections}) and the lower final stellar mass. However, a
sufficiently large mass of stars is formed in the bursts such that the galaxy
still lies several orders of magnitude above the (extrapolated) abundance matching
relations.

\begin{figure*}
\centering
\includegraphics[width=1.0\textwidth]{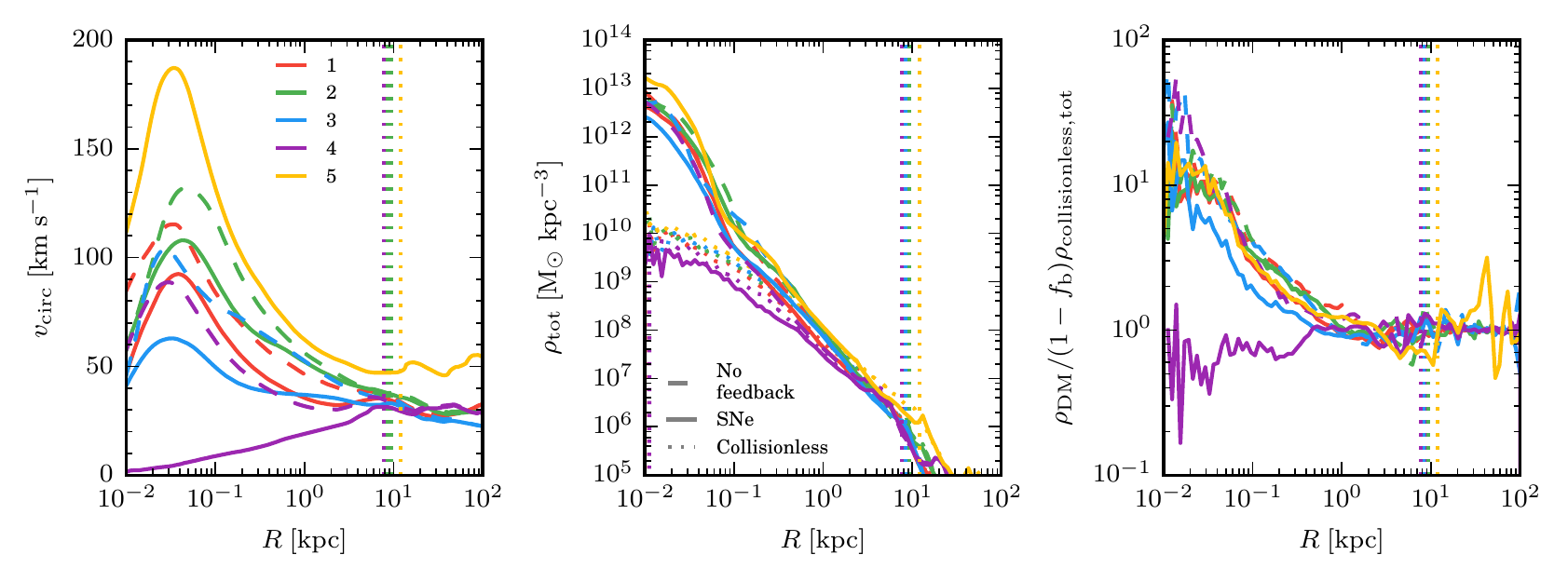}
\caption{Radial profiles at $z=4$ for the various simulations, solid and
  dashed lines indicating runs with and without SNe, respectively. Vertical
  dotted lines denote the virial radius. Dwarf 5 without feedback is not shown
  as it was halted prior to this redshift, however its $z=4.4$ profiles are
  consistent with the results from the other dwarfs. \textit{Left}: circular
  velocity profiles. While SN feedback systematically reduces the peak of
  circular velocity profiles, this reduction is only moderate (with the
  exception of dwarf 4). \textit{Centre}: total (i.e. dark matter, gas and
  stars) density profiles. Dotted lines show profiles from collisionless dark
  matter only simulations. \textit{Right}: the ratio of dark matter density in
  the simulations with baryonic physics as compared to the collisionless
  simulations (renormalised by the cosmic dark matter fraction).}  
\label{structure} 
\end{figure*}

As mentioned previously, despite ending up with a $z=4$ halo mass similar to
the other dwarfs simulated, dwarf 4 spends most of its history as two lower
mass systems prior to a late major merger. Correspondingly, in the runs
without feedback, the progenitors have lower SFRs than the other dwarfs,
although this results in similar stellar to halo mass ratios (see
Fig.~\ref{mhalo_star}). Peaking at $0.05\ \mathrm{M_\odot\ yr^{-1}}$
by $z=9$, the SFR of both galaxies evolves in a similar fashion. There is a
slight drop in SFR after $z=9$. The two haloes merge around $z=5.5$ leading to
a rapid increase in star formation. Like dwarfs 2 and 3, with the addition of
feedback, the initial onset of star formation is limited to a short
burst. However, the system is even more efficiently cleared of gas, resulting
in a complete lack of star formation until the merger occurs. Unlike the no
feedback case, this merger is relatively dry so the merger-triggered star
formation burst is severely curtailed. 

Dwarf 5 starts forming stars at $z\approx15$, rising to high SFR after
$z=10$. A large amount of variability can be seen, mainly corresponding to
mergers. Feedback has very little impact on the SFR in an averaged sense,
although it impacts the gas near the very centre of the halo enough to cause
variations relative to the no feedback simulation.

Fig.~\ref{structure} shows circular velocity profiles, total density profiles
and the ratio of dark matter density in simulations with baryonic physics
compared to collisionless (i.e. dark matter only) simulations at $z=4$. It can
be seen that on the whole, the simulations give rise to extremely concentrated
mass distributions. The circular velocity profiles are strongly peaked at very
small radii (10s of parsecs), in some cases $>100\ \mathrm{km\ s^{-1}}$. The
inclusion of SNe reduces the magnitudes of the peaks by a factor of a
few. Dwarf 4, which has managed to significantly suppress star formation (as
seen in Figs.~\ref{mhalo_star} and \ref{sfr}), is unique in preventing a
peaked circular velocity profile. Instead, a gently rising profile reaches its
peak value $\sim30\ \mathrm{km\ s^{-1}}$ near the virial radius  where it
converges with the no feedback profile. 

The centrally concentrated mass distribution that gives rise to these strongly
peaked circular velocity profiles can be seen in the central panel of
Fig.~\ref{structure} in the form of radial profiles of total density
(i.e. dark matter, gas and stars). Also plotted are profiles from
collisionless simulations (dotted lines). These latter profiles are well fit
by NFW profiles \citep{Navarro1997}. The introduction of baryons leads to a
strong peak of gas and stars with $0.1\ \mathrm{kpc}$, overdense relative to
the collisionless simulations by a factor of 100 in the centre. While the
baryonic mass is dominant in this region, it can be seen in the rightmost
panel of Fig.~\ref{structure} that dark matter density has also been enhanced
by a factor of $\sim10$. Here, we plot the ratio of the dark matter density to
the density from the collisionless simulations (renormalised by the cosmic
dark matter fraction). The central concentration of baryons has lead to
contraction of the dark matter. Only in dwarf 4 has the feedback managed to
expel sufficient baryons to prevent this central overdensity, its total and
dark matter density profiles lying marginally under the collisionless case.

\begin{figure*}
\centering
\includegraphics[width=1.0\textwidth]{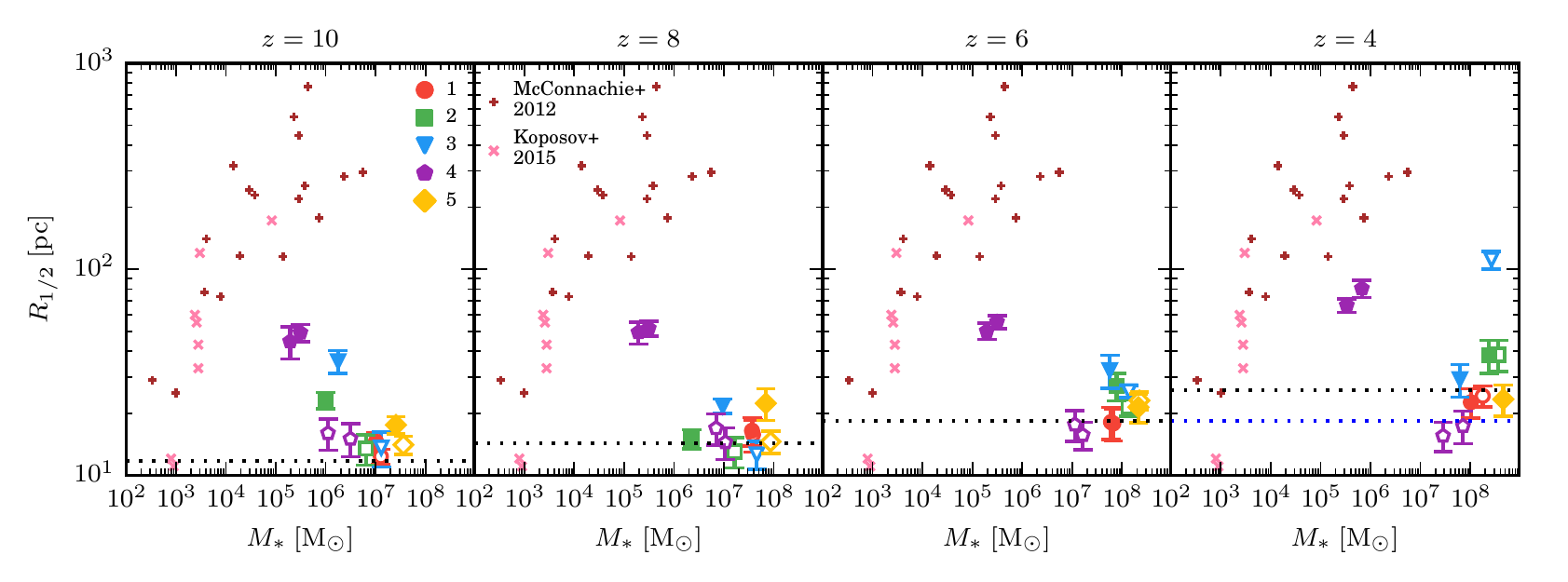}
\caption{Projected stellar half-mass radius vs. stellar mass for the central galaxies at $z=10$, 8, 6 and 4. Once again, in the case of dwarf 4, the central galaxies of the two progenitor haloes are shown. The points are
calculated as the mean over a sample of 500 random viewing angles, with the
error bars marking one standard deviation (open symbols are for no feedback runs,
while filled symbols are for simulations with SNe). Horizontal dotted lines
indicate the gravitational softening length at a given redshift (at $z=4$, the
blue dotted line corresponds to dwarfs 3, 4 and 5). Also plotted are
observations for local dwarfs \citep{McConnachie2012,Koposov2015} for
comparison, although 
a comparison of these $z=0$ objects with our $z=4$ galaxies should be treated with caution. Most objects form most of their stars in a dense central region limited only by the softening length (dwarf 3 no feedback is an outlier, see text). SNe have little impact, except in dwarf 4.}  
\label{sizes} 
\end{figure*}

\begin{figure*}
\centering
\includegraphics{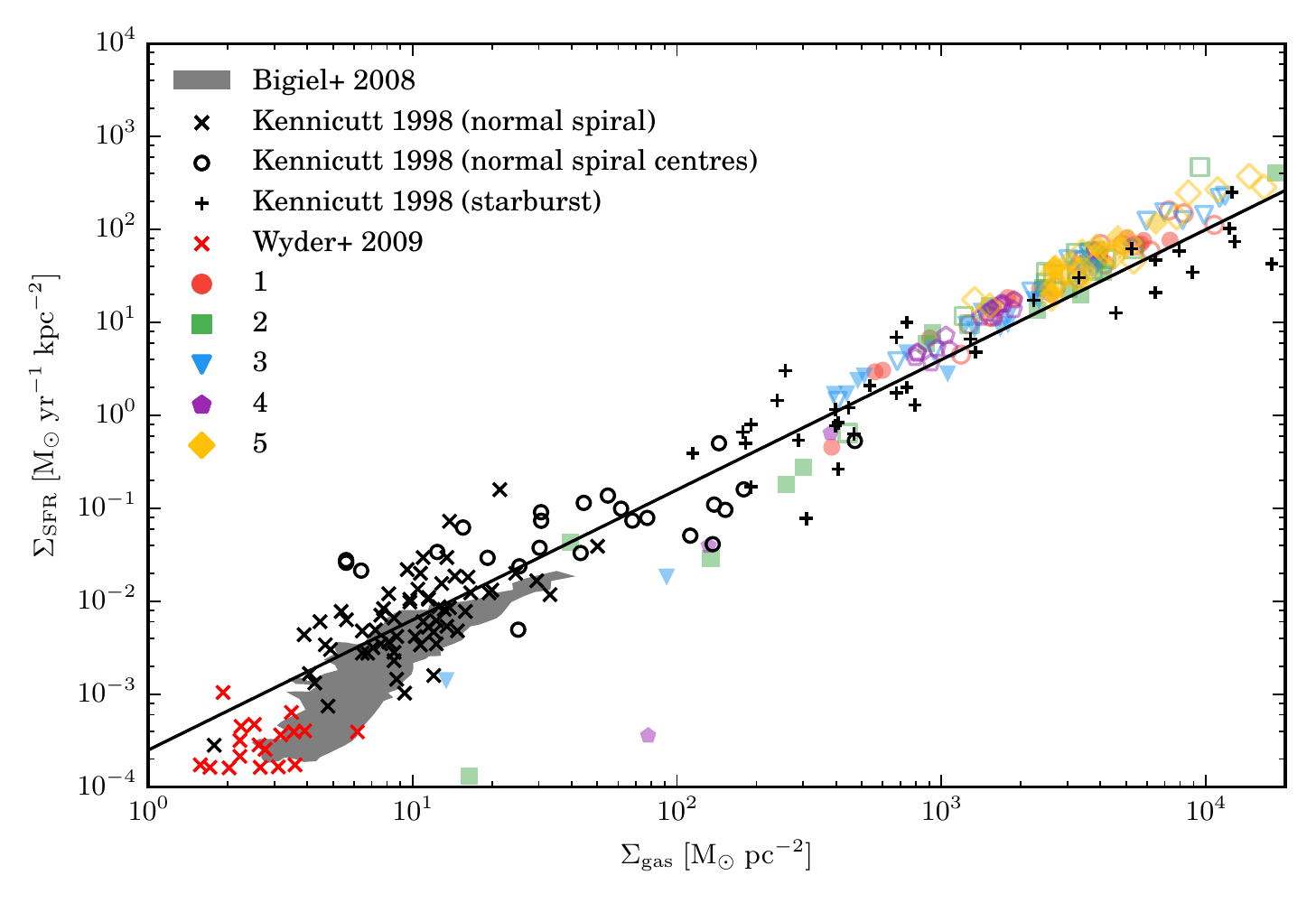}
\caption{A Kennicutt-Schmidt plot, showing SFR surface density as a function
  of gas surface density for our simulated dwarfs between $z=12-4$ ($z=12-4.4$
  for dwarf 5 with no feedback). We only plot the most massive of the two
  dwarf 4 subhaloes for clarity, but the secondary subhalo exhibits similar
  behaviour. These are global measurements, taken within a radius containing
  90\% of the total SFR, projecting down the gas angular momentum vector (open
  symbols are for no feedback runs, 
while filled symbols are for simulations with SNe). Also
  shown are observations, both global \protect\citep{Kennicutt1998,Wyder2009}
  and spatially resolved \protect\citep{Bigiel2008}. We also plot the power
  law fit of \protect\cite{Kennicutt1998} to the data of that work. Most of
  our simulated galaxies have high gas surface densities and SFR surface
  densities. A few galaxies experience strong bursts of feedback which drive
  them well beyond the boundaries of the plot as they are quenched, the few
  low surface density points representing transitions.}   
\label{ks} 
\end{figure*}

\begin{figure*}
\centering
\includegraphics[width=1.0\textwidth]{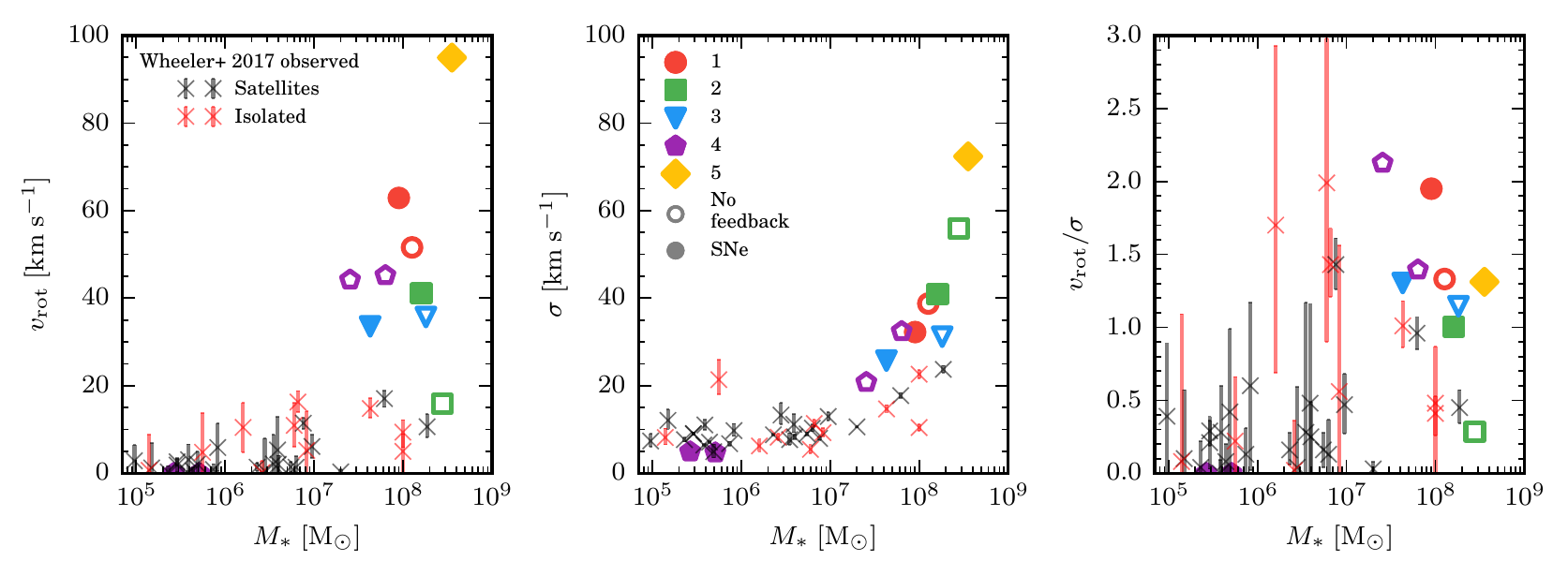}
\caption{Stellar kinematic properties of the simulated dwarf galaxies at
  $z=4$. \textit{Left}: the peak rotational velocity, having aligned the
  system in the `disc' plane relative to the total stellar angular momentum
  vector. \textit{Centre}: the 1D velocity dispersion within the peak
  rotational velocity radius (or in the case of dwarf 4 with SNe, within
  stellar half-mass radius, see text for details). \textit{Right}: the ratio
  of rotational velocity to velocity dispersion, a measure of rotational
  support. Also plotted are measurements of observations from
  \protect\cite{Wheeler2017} of dwarfs in the Local Group.}  
\label{v_sigma} 
\end{figure*}

Fig.~\ref{sizes} shows the 2D projected stellar half-mass radius, $R_{1/2}$,
as a function of stellar mass for the various galaxies at $z=10$, 8, 6 and
4. We make this measurement from 500 randomly distributed viewing
angles\footnote{The results are well converged with number of samples above
  500.}. The mean of the sample is plotted, error bars indicating the 1$\sigma$
limit of the distribution. We mark with horizontal dotted lines the
gravitational softening lengths. For reference, we also plot observations of
local dwarfs \citep{McConnachie2012,Koposov2015}, although the comparison of
these $z=0$ objects with our $z=4$ dwarfs should be taken with some
caution. The majority of our galaxies have extremely compact stellar
distributions. While the projections in Fig.~\ref{projections} show extended
discs on the scale of hundreds of parsecs, most of the stellar mass is
contained within a few tens of parsecs. In fact, the stellar half-mass radius
is very close to the gravitational softening length, indicating that that the
objects have undergone catastrophic collapse halted only by our limited
resolution\footnote{Dwarf 3 without feedback at $z=4$ is somewhat of an
  outlier. The lack of any disruption from mergers has allowed a more extended
  stellar distribution. As the majority of these stars are formed between
  $z=6-4$, this leads to a sudden increase in $R_{1/2}$ by $z=4$. The
  simulation with SNe has a significantly smaller $R_{1/2}$, but this is
  mainly because it has a proportionally lower mass of stars.}. The two
component subhaloes of dwarf 4 remain less concentrated with the inclusion of
SNe, lying at $z=4$ within a factor of a few of the $z=0$ observations at (dwarfs
2 and 3 also have larger $R_{1/2}$ at $z=10$ before the failure of the SNe at
later times).

Fig.~\ref{ks} shows the location of our objects on a Kennicutt-Schmidt plot
(SFR surface density as a function of gas surface density) between
$z=12-4$. We make these global measurements by taking the face-on `disc'
projection defined by the total angular momentum vector of the gas within
twice the 3D stellar half-mass radius (although it should be noted that not
all of our galaxies produce discs). For a given projection of the galaxy, we
find the 2D radius containing 90\% of the total SFR\footnote{The measurements
  are relatively insensitive to the exact fraction adopted, the points being
  shifted up and down the Kennicutt-Schmidt relation slightly.}. We then
compute SFR surface density and mass surface density from the gas within this
radius. Fig.~\ref{ks} also shows global \citep{Kennicutt1998,Wyder2009} and
spatially resolved \citep{Bigiel2008} observations. Due to the extremely
compact nature of most of our galaxies, the majority of our simulations appear
in the same region of the Kennicutt-Schmidt plot as starburst galaxies. There
is a trend for our simulations with feedback to produce galaxies with slightly
lower SFR and mass surface densities than simulations without feedback. When
galaxies experience an efficient burst of feedback (dwarfs 2, 3 and 4; see
Fig.~\ref{sfr}) they move towards the lower end of the relation. However,
because these bursts are very strong and tend to completely disrupt the star
forming gas, we do not see a steady state at low surface densities, but the
measurements in this quenched phase lie well beyond the boundaries of the
plot. For example, dwarf 4 with feedback only appears on the plot at $z=11.5$,
11 and 4.5 because it effectively has no star formation at other times (we do
not plot the secondary subhalo of dwarf 4). 

Fig.~\ref{v_sigma} shows kinematic information of our simulated galaxies as a
function of stellar mass at $z=4$ as compared to measurements of local dwarfs
from \cite{Wheeler2017}. The left panel shows the rotational velocity. We take
here the peak value of the stellar rotation curve, having first transformed
into the `disc' plane of the galaxy by aligning with the total angular
momentum vector of the stars. It should be noted that the kinematics from the
simulations should be treated with caution given that the size of the systems
approaches the gravitational softening length in those cases in which
catastrophic collapse has occurred. The rotational velocities are well in
excess of the observations, but not unexpected given the highly peaked
circular velocity profiles (see Fig.~\ref{structure}). It should,
however, again be noted that we are comparing high redshift kinematics with 
low redshift data; we would expect the circular velocity to scale as $(1+z)^{1/3}$
\citep[e.g.][]{Bullock2001} which might reduce the tension. There is a trend for
the simulations with SN feedback to produce higher rotational velocity systems
(except for dwarf 4). With SNe, however, the two subhaloes of dwarf 4 show no
evidence of rotation and are therefore consistent with the observations at
that mass which demonstrate little or no rotation. 

The central panel of Fig.~\ref{v_sigma} shows the 1D velocity dispersion,
$\sigma$, for our systems. We measure the 3D velocity dispersion within a
sphere whose radius corresponds to the peak rotational
velocity\footnote{Taking other reasonable radii, such as one or two times the
  stellar half-mass radius yields the same results within
  $1\ \mathrm{km\ s^{-1}}$; these radii are all comparable.}, then obtain the
1D value by dividing by $\sqrt{3}$. In the case of dwarf 4 with SNe (which
shows no rotation) we use the stellar half-mass radii. Again, most simulations
lie significantly above the local observations, a consequence of the highly
compact systems (Fig.~\ref{sizes} demonstrates how much more extended observed
galaxies in this mass range are). There is a steep relation of increasing
$\sigma$ with increasing stellar mass. The two subhaloes of dwarf 4 with SNe
lie close to the observations, with velocity dispersions of
$\sim5\ \mathrm{km\ s^{-1}}$. 

Examining the ratio of the rotational velocity to the velocity dispersion
provides a measure of the rotational support of the system. This is shown in
the right panel of Fig.~\ref{v_sigma}. Most of our systems are rotationally
supported, in contrast with the observations which prefer rotation to be
subdominant (although there are a few outliers and the uncertainties are large
in some observations), with the caveat that we are comparing our $z=4$ objects
with local observations. Only dwarf 4 with SNe is consistent with observations
producing a dispersion dominated system. We note that our dwarfs 
end up as either over-massive discs when feedback is inefficient 
or a dim spheroidal when feedback is efficient in dwarf 4 
\citep[a similar pattern is found for Milky Way mass
haloes by][]{Roskar2014}, but given that we only have one of the latter
types of objects we cannot make any claim to bimodality.

\begin{figure}
\centering
\includegraphics{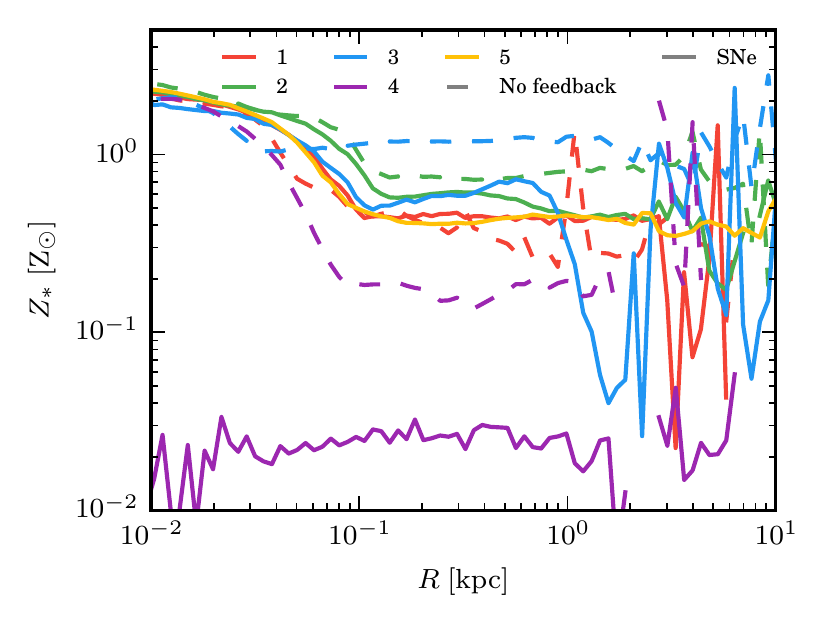}
\caption{Radially (mass-weighted) averaged stellar metallicity profiles at
  $z=4$. Outside of a few kpc, the profiles become very noisy, in some cases
  because of substructures. For dwarf 4, the profiles are centred on the most
  massive subhalo from the no feedback simulations. While in dwarfs 1, 2, 3
  and 5 inefficient SN feedback leads to over-enrichment, in the case of dwarf
  4 this is reduced by two orders of magnitude to more reasonable values.}  
\label{profiles_met}
\vspace{-4ex} 
\end{figure} 

Because dwarfs 1, 2, 3 and 5 produce such large masses of stars in a confined
region, the resulting metal enrichment of the surrounding region is
necessarily extremely high. Fig.~\ref{profiles_met} shows radial stellar
metallicity profiles at $z=4$. Without SNe, the central tens of parsecs (which
contain most of the stellar mass) are dominated by a stellar population of
$\sim2\ \mathrm{Z_\odot}$. The metallicity drops rapidly through the disc
region (on the order of $100$~pc, see also Fig.~\ref{projections}) to reach a
metallicity ranging between $0.2\ \mathrm{Z_\odot}$ (dwarf 4) and
$0.6\ \mathrm{Z_\odot}$ (dwarf 2) in the stellar `halo'. The metallicity
gradient is then flat until the edge of the stellar distribution, after which
the profiles are noisy due to the low stellar density and the presence of
other subhaloes. With the exception of dwarf 4, the addition of SNe makes very
little difference to the stellar metallicities, which is to be expected given
the inefficiency of feedback in these systems (although, in dwarf 3, the
stellar halo is curtailed at a smaller radius). For dwarf 4, SNe reduce the
central metallicities by 2 orders of magnitude to
$0.02-0.03\ \mathrm{Z_\odot}$. The resulting metallicity gradient is flat
through the entire stellar distribution, out to $\sim1\ \mathrm{kpc}$ (the
second subhalo appears in this radial profile at larger radii, as can also be
seen in the no feedback profile). While the lower stellar metallicities are
partially due to the lower overall stellar mass formed relative to the no
feedback simulation, the ability of the SNe to expel metals from the centre of
the halo is also key.

\begin{figure}
\centering
\includegraphics[width=0.4\textwidth,trim=0 8 0 0,clip]{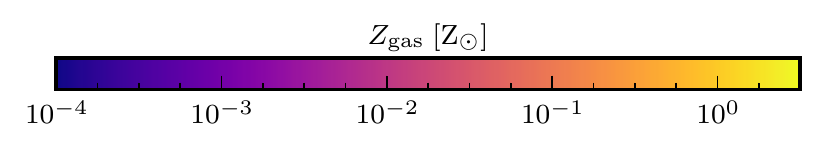}
\includegraphics[trim=0 0 0 6,clip]{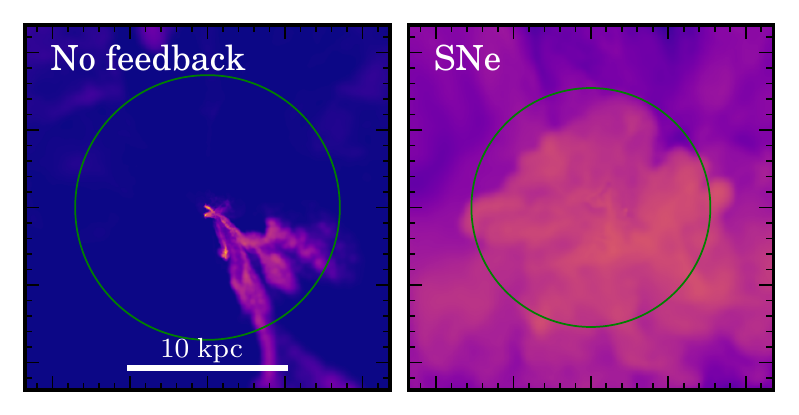}
\caption{Mass-weighted projections of gas metallicity for dwarf 4 at $z=4$,
  without (\textit{left}) and with (\textit{right}) SN feedback.
The virial radius of the halo is marked with a green circle. There is a stark
difference in the gas metallicity distribution, which is much more homogeneous
in the run with SN feedback, allowing the CGM to be enriched to a few times
$10^{-2}\ \mathrm{Z_\odot}$.}
\vspace{-4ex}  
\label{met4} 
\end{figure}

Fig.~\ref{met4} shows mass-weighted gas metallicity projections of dwarf 4,
without and with the inclusion of feedback. In the absence of feedback, metals
remain where they have been deposited by the star particles, leading to high
concentrations around the subhaloes. This can be seen in the projection, where
the majority of gas (both inside and outside of the virial radius) remains at
the metallicity floor of the initial conditions,
$10^{-4}\ \mathrm{Z_\odot}$. Very small patches of super-solar metallicity gas
can be seen around the subhaloes, while some metal enriched gas has been
stripped during the merger, leaving trails. With the inclusion of feedback,
gas of a few $10^{-2}\ \mathrm{Z_\odot}$ is widely distributed inside and
outside of the virial radius. In the other dwarfs, the inclusion of feedback
also allows metals to leave the halo ($\sim10^{-1}\ \mathrm{Z_\odot}$ at the
virial radius), but approximately 3 orders of magnitude more stellar mass has
been created to achieve this i.e. the SNe are $\sim100$ times less efficient
at ejecting metals. We reported a similar phenomenon in isolated simulations
in \cite{Smith2018}, where inefficient SNe lead to slow moving, highly metal
enriched outflows simply due to the number of SNe occurring. Nonetheless, dwarf
4 demonstrates that it is possible for dwarfs to efficiently enrich the CGM.

\begin{figure*}
\centering
\includegraphics[width=1.0\textwidth]{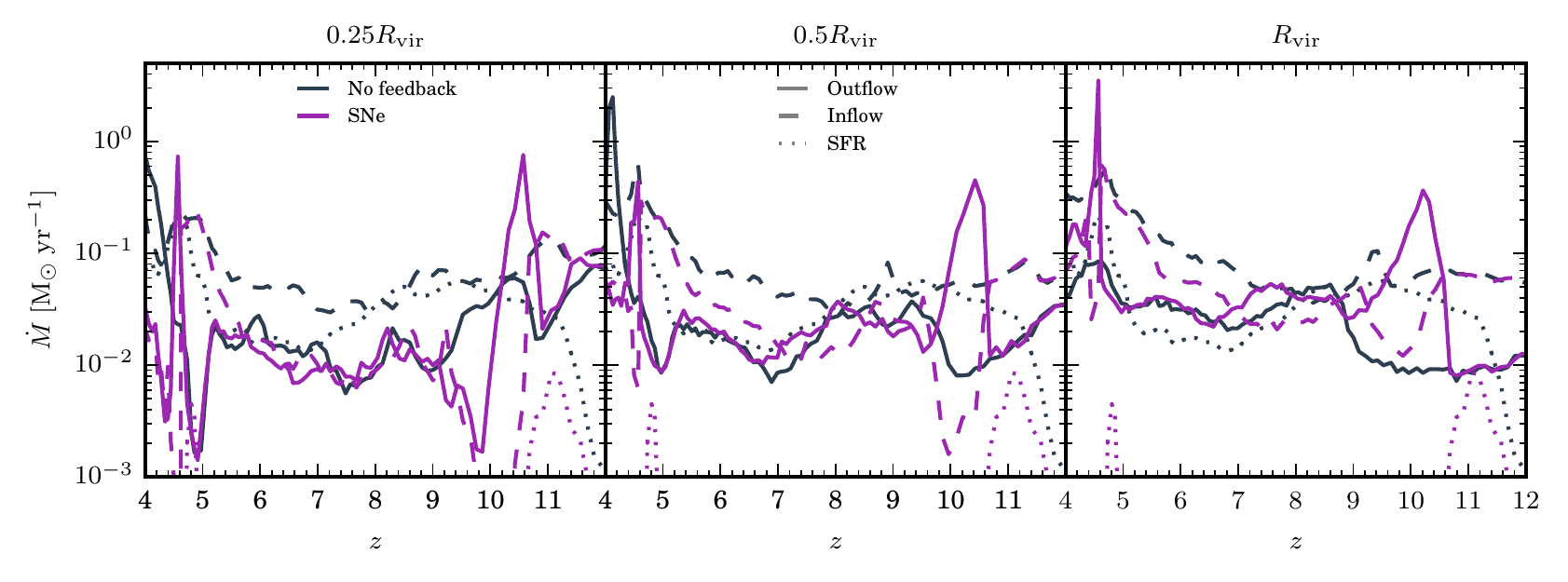}
\caption{Mass outflow and inflow rates as a function of redshift for dwarf 4,
  with and without feedback. Rates are calculated across shells of thickness
  $50$~pc located at $0.25R_\mathrm{vir}$, $0.5R_\mathrm{vir}$ and
  $R_\mathrm{vir}$. The SFR is also shown for reference. With SN feedback, a
  burst of star formation causes a subsequent outflow with mass loading
  factors between $10-100$ depending on the radius (comparing peak SFR to peak
  outflow rate) as well as strongly suppressing inflow. Sudden increases in
  inflow and outflow rates in both feedback and no feedback runs near $z=4$
  are largely due to the merger.}   
\label{flow4}
\vspace{-4ex} 
\end{figure*}

Fig.~\ref{flow4} shows gas mass outflow and inflow rates across
$0.25R_\mathrm{vir}$, $0.5R_\mathrm{vir}$ and $R_\mathrm{vir}$ as a function
of redshift for dwarf 4 with and without SN feedback. SFRs are also plotted for comparison. The outflow rates are calculated as:
\begin{equation}
\dot{M}_\mathrm{out} = \frac{\sum_i m_i v_{\mathrm{out},i}}{\Delta r} \,,
\end{equation}
where the sum is over all gas cells within a shell of thickness $\Delta r =
50\ \mathrm{pc}$ centred on the target radius that have a positive radial
outflow velocity, $v_\mathrm{out}$. The inflow rates are calculated in the
same manner, but for all cells that have a negative radial velocity. In the
absence of SN feedback, outflow rates mostly remain well below inflow rates,
with outflows only arising from the motion of substructures and mergers. For
example, the dramatic increase in outflow rates just before $z=4$ is due to
the motion of the merging subhalo within the primary halo. With SN feedback,
after the bursts of star formation, outflow rates increase dramatically while
inflow is suppressed. 

Outflows are often characterised in terms of mass loading factor, i.e. the
ratio of outflow rate to SFR. The outflows across the three radii are offset
from the peak of the SFR due to the time difference between star formation and SNe
exploding as well as the travel time of the outflow, so an instantaneous mass
loading factor is not a useful quantity. However, comparing the peak SFRs and
outflow rates yields a mass loading factor of approximately 90, 60 and 30
across $0.25R_\mathrm{vir}$, $0.5R_\mathrm{vir}$ and $R_\mathrm{vir}$,
respectively. Following the burst of star formation at $z=11$, inflow across
$0.25R_\mathrm{vir}$ is essentially halted until after $z=10$. The inflow
rates remain a factor of $\sim5$ below the corresponding no feedback
simulation rates until the merger begins at $z\approx5.5$. At this point, it
appears that the UV background is hindering the ability of gas to condense
into the centre of the halo.  
The second burst of star formation after $z=5$ also produces a brief outflow,
preventing further star formation. None of the other dwarfs simulated are able
to produce strong outflows. Dwarfs 2 and 3 have very brief outflows after
bursts of star formation and subsequent efficient feedback, but they have mass
loading factors $<2$ and barely suppress inflow rates except in the very
centre of the halo. 

\vspace{-4ex}
\section{Discussion} \label{zDiscussion}
\begin{figure*}
\centering
\includegraphics{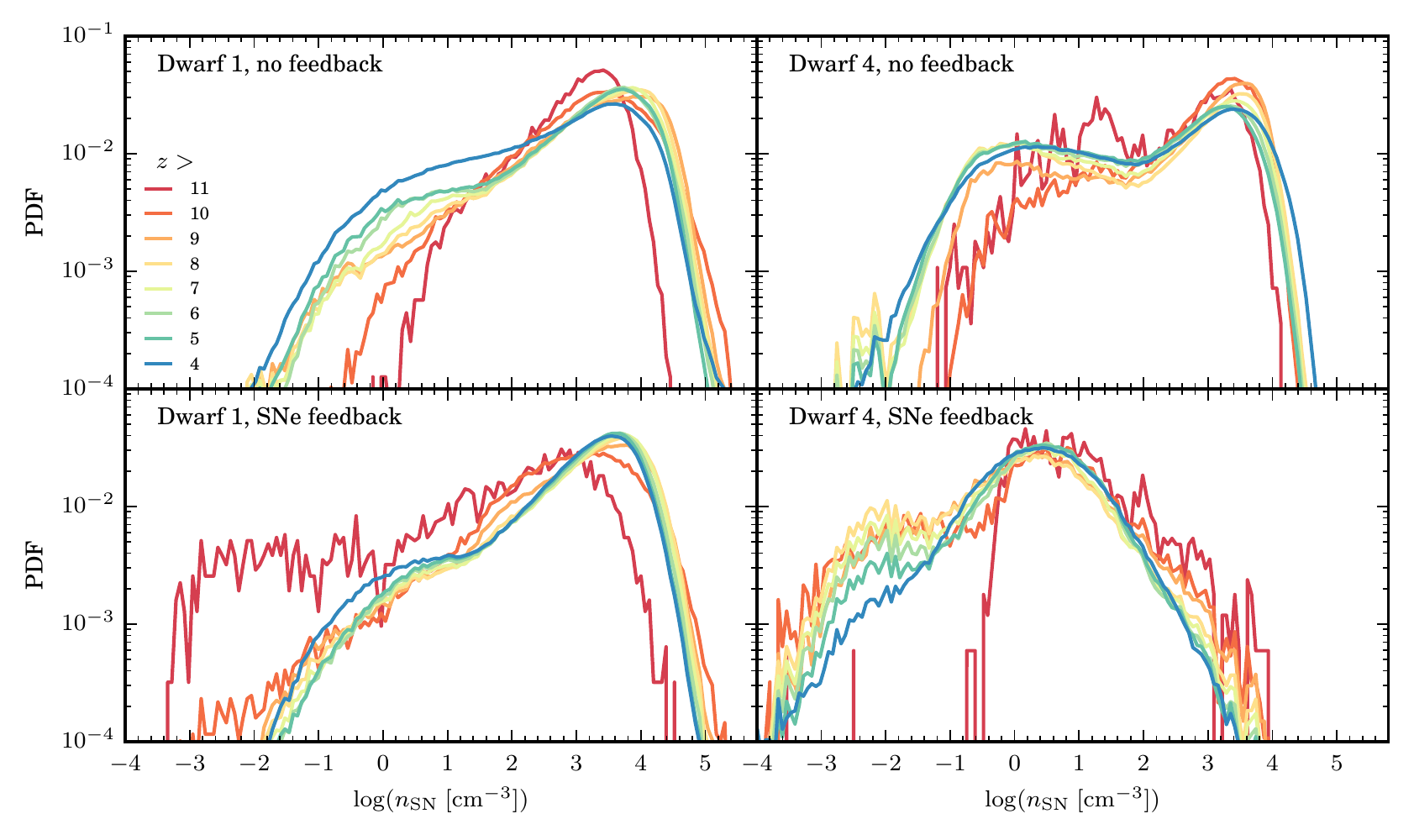}
\caption{Distribution of the densities of gas in which SNe occur. Dwarf 1 and
  4 are compared, with and without feedback. The redshift evolution of the
  PDFs are shown (cumulatively). For numerical reasons, these PDFs are for all
  SNe that occur in the high-resolution region, rather than being tied
  explicitly to the host halo of a given dwarf. However, the vast majority of
  SNe occur in the host halo, so these PDFs are representative. Most SNe occur
  in gas with a density of approximately $10^4\ \mathrm{cm}^{-3}$ for dwarf 1
  (with and without feedback) and dwarf 4 without feedback. However, with the
  inclusion of SN feedback in dwarf 4, the mean density drops by three orders
  of magnitude.} 
\label{dens_pdf} 
\end{figure*}

\subsection{Why is SN feedback inefficient?}
In all of our simulated galaxies except for dwarf 4, SN
feedback is unable to prevent the catastrophic collapse of gas and resulting
runaway star formation. The reason for this inefficiency appears to be that
most SNe occur in very dense gas. This can be seen in Fig.~\ref{dens_pdf} which
shows the distribution of gas density in which SN explode for all SNe above a
given redshift for dwarfs 1 and 4, with and without feedback. Comparing the
PDFs for dwarf 1 at $z=11$ for the no feedback and feedback simulations, both
peak at a high density of $\approx10^{3}\ \mathrm{cm^{-3}}$. With feedback,
there is a slight tail to low density, indicating that the feedback has been
able to clear some gas. A short time later at $z=10$, the peak of the
distribution is at $\approx10^{4}\ \mathrm{cm^{-3}}$. By contrast, while
without feedback dwarf 4 has a similar PDF to dwarf 1, once SNe are included
the peak of the distribution is at $\sim3\ \mathrm{cm^{-3}}$ for all
redshifts. With SNe as the sole form of feedback, the decisive 
criterion determining the success and failure of the feedback is whether it is 
able to clear the dense gas immediately. If at any point it cannot, 
then subsequent SNe will become increasingly inefficient, eventually resulting in the
inability of the feedback to have sufficient impact on galaxy properties. 

Dwarf 1 fails the criterion
immediately, as does dwarf 5. Dwarf 2 succeeds twice but is overwhelmed by a
sudden increase of gas during a wet merger at $z\sim6$. Dwarf 3 is partially
successful, but the bursts reach too high a SFR before the system is quenched,
so the net reduction in stellar mass is too low. Dwarf 4 is unique amongst our
simulations in being completely successful, mainly due to its merger
history. As described in the previous section, while the final halo is
comparable in mass to dwarf 1, it is formed from a major merger (with a mass
ratio $\sim1.5$) late in its history ($z\approx5.5$). This means that it
spends most of its evolution as two smaller haloes. Having a lower halo
mass makes it easier to clear gas for two reasons. Firstly, there is a shallower 
potential well to fight against. Secondly, the inflow of gas onto the haloes is 
reduced relative to dwarf 1 (even without feedback, the SFRs for the two progenitor 
haloes are lower than for dwarf 1; see Fig.~\ref{sfr}). This means
that while at $z=4$ dwarf 1 and dwarf 4 have the same virial mass to within 0.05~dex,
they have evolved as if they were very different mass systems due to the manner
of their assembly.

It may appear at first glance that the inefficiency of SNe in dense gas is a
result of shortcomings in our method of feedback injection i.e. numerical
overcooling. However, our mechanical scheme is designed to help 
mitigate the effects of under-resolved SN remnants by injecting the correct 
momentum relative
to the stage of their evolution that can be resolved. Full details can be
found in \cite{Smith2018} where we also demonstrate using isolated simulations
that this scheme is numerically robust (see also the following section where
we discuss convergence with resolution). For extremely dense gas, at most
tractable resolutions, the SN remnant will remain entirely unresolved so our scheme will inject the final momentum achieved during the Sedov-Taylor phase. We make use of a fitting function to high resolution simulations of individual SNe \citep[see][]{Blondin1998, Thornton1998, Geen2015, Kim2015,Martizzi2015,Kimm2015},
\begin{equation}
p_\mathrm{fin} = 3 \times 10^5\ \mathrm{km\ s^{-1}\ \mathrm{M}_{\rm \odot}}\,
E^{16/17}_{51} n^{-2/17}_{\mathrm{SN}} Z^{-0.14}_\mathrm{SN}\,, 
\label{p_fin}
\end{equation}
where $E_{51} = \left(E_\mathrm{SN} / 10^{51}\ \mathrm{ergs}\right)$ is the
energy of the SN (for our individually time-resolved SNe,
$E_\mathrm{51}\equiv1$), while $n_\mathrm{SN}= \left(n_\mathrm{H} /
\mathrm{cm^{-3}}\right)$ and $Z_\mathrm{SN} = \mathrm{MAX}\left( Z/Z_{\rm
  \odot}, 0.01\right)$ are the hydrogen number density and metallicity of the
ambient gas, respectively. 

It can therefore be seen by comparing the peaks of the density PDF for SN
sites that in dwarf 1 the momentum budget per SN is reduced to $\sim0.39$ of
that in dwarf 4. Additionally, if metals are not cleared efficiently this will
also impact the available momentum. Given that the typical gas metallicity in the
centre of dwarf 1 is approximately a factor of 100 higher than in dwarf 4,
this reduces the momentum budget again by half, meaning that in total only
20\% of the momentum budget per SN is available relative to dwarf 4. In addition
to impacting the small scale evolution of the SN remnant, the build up of a
central concentration of dense gas will make it more difficult for the
momentum injection from SNe to clear material from the galaxy because the mass
of material that must be swept up in order for an outflow to escape becomes
proportionally higher. These two factors lead to a state of runaway star
formation if at any point the feedback is unable to prevent the build up of
dense gas, particularly if inflow rates increase suddenly (e.g. due to
mergers, be they major or minor). 

\vspace{-2ex}
\subsection{The impact of the choice of parameters on our results}\label{Section_parameters}
\begin{figure*}
\centering
\includegraphics{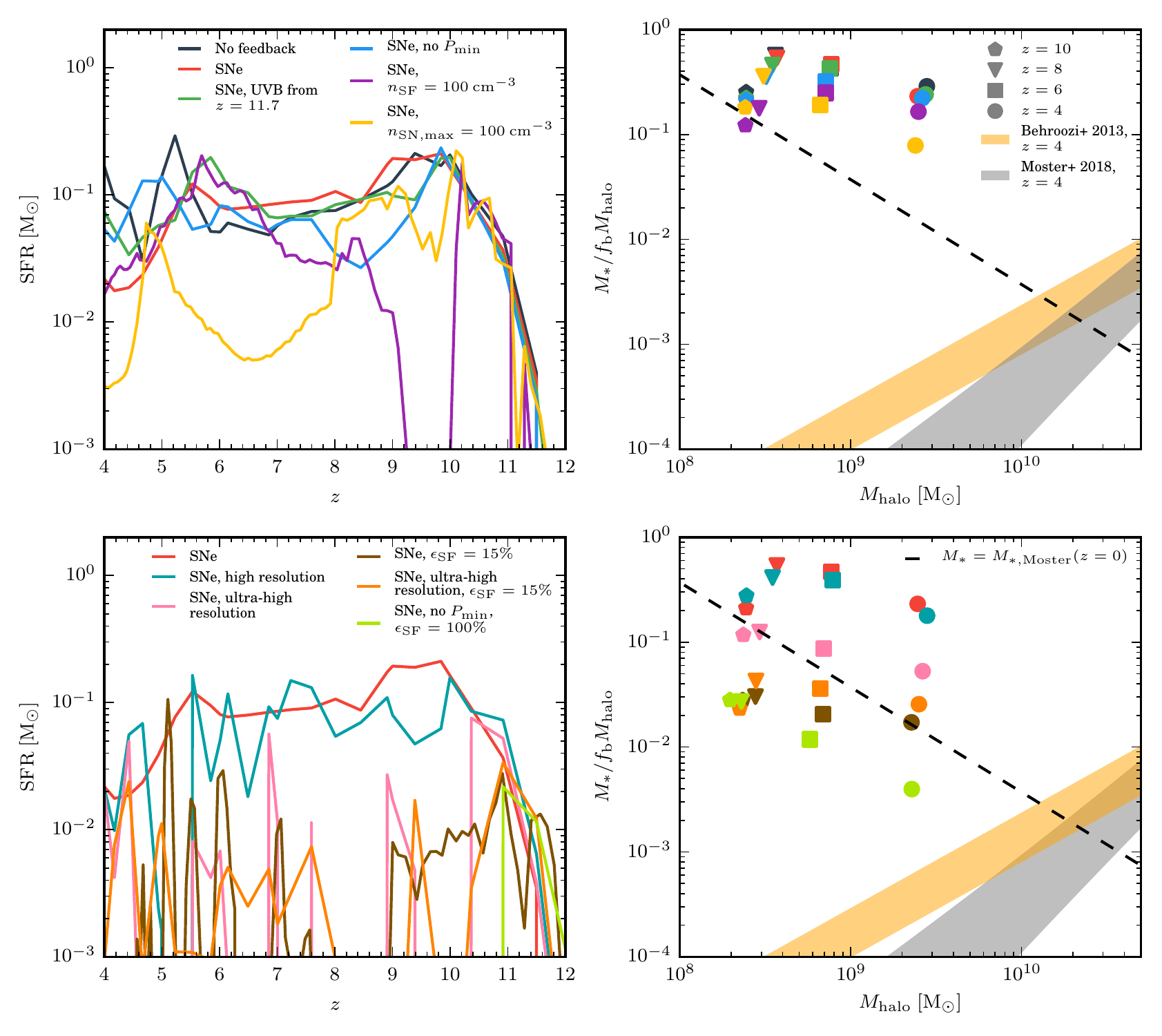}
\caption{SFR as a function of redshift (\textit{left}) and integrated star
  formation efficiency (\textit{right}) for simulations of dwarf 1 with a
  variety of alternative parameters. The results are split into two rows for
  clarity. Simulations are at the fiducial 
  gas cell mass resolution of 287~$\mathrm{M_\odot}$ unless denoted
  high resolution (35.9~$\mathrm{M_\odot}$) or ultra-high resolution (15~$\mathrm{M_\odot}$).
  The simulations shown are as follows.
  \textit{Top}: no feedback (black), our fiducial SNe simulation (red), the
  \protect\cite{FG2009} UV background is turned on from our first available
  tabulated redshift of $z=11.7$ (green), the pressure floor is turned off (blue),
  the star formation density threshold is increased by a factor of 10 to $100\ \mathrm{cm^{-3}}$
  (purple),
  we impose a cap of $100\ \mathrm{cm^{-3}}$ on the density that is used to determine the maximum
  momentum that can be injected for a SN (see eq.~\ref{p_fin}) (yellow).
  \textit{Top}: the fiducial SNe simulation is repeated in these panels for reference (red),
  high resolution (cyan), ultra-high resolution (pink),
  fiducial resolution with the star formation efficiency increased by a factor of 10 to 15\%
  (brown),
  ultra-high resolution with the star formation efficiency increased to 15\% (orange),
  fiducial resolution with the pressure floor is turned off and the star formation efficiency
  is set to 100\% (light green). Abundance matching relations at $z=4$
  \protect\citep{Behroozi2013,Moster2018} are shown, although they are extrapolated
  into this mass range. The dashed black line indicates the integrated star formation efficiency
  at a given halo mass if the stellar mass equalled that predicted from the $z=0$ halo mass
  by the \protect\cite{Moster2018} relation (still slightly extrapolated to this mass, even at $z=0$). If a simulation exceeds this line at any point, the galaxy
  will overshoot the $z=0$ abundance matching relation even if it does not form any more stars.}
\label{parameters} 
\end{figure*}

Having discussed the reasons why SN feedback is inefficient in our fiducial
simulations, we now explore the degree to which our results are generally
applicable as opposed to being dependent on our choice of
parameters. Fig.~\ref{parameters} shows the SFR as a function of redshift and
the stellar mass to halo mass ratio as a function of halo mass (at $z=10$, 8,
6 and 4) for our fiducial simulations of dwarf 1 as well as 9 resimulations in
which we vary various parameters of our models. Again, we plot the (heavily extrapolated) $z=4$ abundance
matching relations from \cite{Behroozi2013} and \cite{Moster2018}. We also
indicate the integrated star formation efficiency equivalent to the stellar mass equalling
the (still extrapolated) \cite{Moster2018} $z=0$ prediction\footnote{We take the halo mass at $z=0$ for dwarf 1 from a
dark matter only simulation. Using the abundance matching relation from \cite{Moster2018} we obtain a predicted
stellar mass, $M_{*,\mathrm{Moster}}(z=0)$. Even at this redshift, we must still extrapolate down in halo mass by 0.5 dex.
We can then determine the equivalent integrated star formation efficiency for a given halo mass (i.e. at a higher redshift)
if the galaxy had the predicted $z=0$ stellar mass. This is the dashed line in Fig.~\ref{parameters}.}. 
If the galaxy exceeds this value, this means that it has already formed more than the $z=0$ stellar mass prediction (although this should be taken as a rough guide because of the effects of extrapolation and intrinsic scatter). It can be seen that this is the case for the majority
of our simulations, often by $z=10$.

We can see that our choice to delay turning on the \cite{FG2009}
UV background until $z=9$ is similar to switching it from $z=11.7$, apart from
a slight reduction in SFRs between $z=10-9$. This shows that the assumed UV
background is unable to prevent the catastrophic build up of gas at $z=10$. We
note, however, that this conclusion rests on the approximation of a
homogeneous UV background as opposed to local varying radiation fields.
Dwarfs that are in crowded regions or are satellites of larger galaxies may be
bathed in ionizing radiation from nearby external sources, assuming that those
galaxies are able to clear/ionize sufficient local gas to achieve a high
enough escape fraction for UV photons. The failure of the UV background
to quench our dwarfs is not inconsistent with other works that
indicate the existence of a $z=0$ threshold mass of a few
$10^9\ \mathrm{M_\odot}$ below which UV background is effective
\cite[see e.g.][]{Okamoto2008, Shen2014, Sawala2015, Wheeler2015, Fitts2017}
as our dwarfs will have 
$z=0$ virial masses in excess of $10^{10}\ \mathrm{M_\odot}$. 
Perhaps more importantly, we have
neglected photoionization from the stars formed in the galaxies
themselves. This may be able to prevent the build up of dense gas in star
forming regions \citep[see
  e.g.][]{Vazquez-Semadeni2010,Walch2012,Dale2014,Sales2014,Rosdahl2015}.
The formation of H\,\textsc{ii} regions can result in SNe occurring
in lower density regions, enhancing their efficiency \citep[see e.g.][]{Geen2015}.
However, resolving H\,\textsc{ii} regions in these circumstances is challenging 
(the radius of a Str{\"o}mgren sphere around a typical O star embedded in $10^4\,\mathrm{cm^{-3}}$ gas is sub-parsec). While it is possible to try and compensate for the
effect of unresolved H\,\textsc{ii} regions on SN remnant evolution in a subgrid
manner \citep[e.g.][]{Kimm2017}, this is beyond the scope of this work. However,
we note that pre-SN stellar feedback is one example of additional physics 
that can enhance the ability of SNe to regulate galaxy properties, as we 
discuss below.

The use of pressure floors to prevent artificial fragmentation is a subject of
some debate in the literature. We discussed the impact of adopting such a
technique in some detail in \cite{Smith2018}, so we refrain from an in-depth
discussion here. Nonetheless, we tested the impact on our zoom-in simulations
by resimulating dwarf 1 without a pressure floor. As can be seen from
Fig.~\ref{parameters}, this has a negligible impact on our results. The lack
of a floor seems to produce slightly more clustered SNe, leading to a
reduction in SFR by a factor a few from $z\sim10-7$. However, in general the
SFR is similar to the fiducial simulation and the $z=4$ stellar mass is the
same within 4\%.

Increasing the star formation threshold by an order of magnitude to
$100\ \mathrm{cm^{-3}}$ also produces more clustered SNe at early times,
allowing the feedback to quench star formation at $z=10$. This leads to a
reduction in stellar mass relative to the fiducial case by a factor of a few
at $z=8$. However, this is still not enough to prevent the build up of dense
gas at later times. From $z=7$ onwards, the SFR is similar to the fiducial
case, leading to a reduction of the $z=4$ stellar mass by only 1.3.

We further carry out a simulation in which we modify the equation used to
calculate the final momentum of a SN remnant after the Sedov-Taylor phase
(eq.~\ref{p_fin}) such that the dependence on ambient gas density is capped at
$100\ \mathrm{cm^{-3}}$. This is a crude approximation to the idea that local
stellar feedback may have prevented surrounding gas reaching high density
prior to the first SN occurring. Imposing this density cap increases the
momentum budget per SN by a factor of 1.7 relative to SN occurring in
gas with $10^{4}\ \mathrm{cm^{-3}}$ (as in Fig.~\ref{dens_pdf}). Of course, the gas
itself is still at high density, so continues to present an obstacle to efficient
clearing of material from a hydrodynamical standpoint. Nonetheless, with this
caveat in mind and very moderate increase in the momentum budget, this
simulation results in a factor of 3 lower stellar mass at $z=4$. This hints
that the need to regulate local gas density is important, but also
demonstrates that such a simple modification to the subgrid scheme is not
sufficient to obtain realistic galaxy properties.

Increasing the number of resolution elements in the zoom-in region by a factor
of $2^3$ (which we label as `high resolution'), giving a mass resolution of $191\ \mathrm{M_\odot}$ and
$35.9\ \mathrm{M_\odot}$ for dark matter particles and gas cells,
respectively, has very little impact on the results. While the SFR shows
slightly more variation than the fiducial resolution simulation and there is a
suppression of star formation briefly between $z=5.5-5$, even with the
increased resolution the feedback is unable to prevent runaway star formation
beginning at early times. This leads to a $z=4$ stellar mass that only differs
from the fiducial simulation by a factor of $1.2$. We also further increase
the gas cell mass resolution to $15\ \mathrm{M_\odot}$ (labelled `ultra-high resolution')\footnote{Our refinement/derefinement
strategy keeps the cells within a factor of 2 of the target mass. At $15\ \mathrm{M_\odot}$ this means that a 
substantial number of star particles are formed with an initial mass lower than
our fiducial SN ejecta mass of $10\ \mathrm{M_\odot}$. Therefore, for this simulation
we drop the ejecta mass to $5\ \mathrm{M_\odot}$. We have carried out additional
tests (not shown) to confirm that this has a negligible impact on our results.}. This results in a far more bursty SFR as outflows
are stronger. This is consistent with our previous tests in isolated setups
that demonstrate that strong outflow generation is difficult to achieve with a
mass resolution coarser than $20\ \mathrm{M_\odot}$ (\citealt{Smith2018}, see also discussions in \citealt{Kimm2015}, \citealt{Hu2016} and \citealt{Hu2019}). While this
means that the results do not converge well with resolution, the galaxy
still exceeds the predicted $z=0$ stellar mass as early as $z=8$. 

At the fiducial resolution, increasing
the star formation efficiency, $\epsilon_\mathrm{SF}$ by a factor of 10 to
15\% leads to significantly different behaviour. The SFR rises faster and
strong clustering of SNe leads to efficient launching of outflows and the
suppression of star formation. Star formation proceeds in short bursts for the
entire duration of the simulation. Despite this, the $z=4$ stellar mass is
only reduced by slightly over an order of magnitude, leaving it over 2 orders
of magnitude above the (extrapolated) abundance matching relations and an
order of magnitude larger than dwarf 4 with the fiducial star formation
parameters. Failing to match the abundance matching relations at this redshift
is not necessarily a failure in and of itself because of the uncertainties
involved at this mass range. However, at $z=4$, the galaxy has
just reached the predicted $z=0$ stellar mass. Given that there are no indications
that it has been conclusively quenched at $z=4$, this suggests that the galaxy
may well end up with an unphysically large stellar mass at $z=0$. Repeating this
experiment with the ultra-high resolution (decreasing cell mass by a factor of
19) reveals similar results, actually resulting in a slightly higher final
stellar mass.

Finally, we try an extreme choice of parameters in an attempt to reduce the
stellar mass further. We turn off the pressure floor and use
$\epsilon_\mathrm{SF}=100\%$. This leads to extremely rapid star formation and
a concentrated burst of SN feedback that is able to completely quench the
galaxy, expelling most of the gas. Star formation does not resume by
$z=4$. The result is a reduction in $z=4$ stellar mass by almost 2 orders of
magnitude. While this is still too high relative to the extrapolated abundance
matching relations, it is possible that this galaxy would move onto the
relation at lower redshift. While this may be seen as a successful solution, 
a more cautious interpretation would indicate
that, given that we need to push our star formation model to its extremes
in order to be successful, we are likely neglecting some other important physical processes
that would alleviate the need for very high values of $\epsilon_\mathrm{SF}$
in the first place. 

Selecting the appropriate value of $\epsilon_\mathrm{SF}$ to use
in galaxy simulations is non-trivial, particularly in the case where
star forming regions may be partially resolved. It is important to establish
over what time and length scales the efficiency is averaged and the degree
to which these scales are relevant to the scales and physics resolvable in the simulation.
We have adopted a fiducial value of $1.5\%$, which represents an average over a
GMC and over a cloud-scale free-fall time \citep[see e.g.][]{Krumholz2007}. 
The value itself can be considered representative for a `typical' Milky Way (MW) star forming region,
although observations reveal a large observed scatter of up to 0.8 dex 
\citep[see e.g.][]{Murray2011,Lee2016}. One way of explaining this scatter is to invoke
a (magneto)-turbulent model of GMC star formation to modulate the efficiency
per free-fall time \citep[e.g.][]{Krumholz2005,Padoan2011,Hennebelle2011,Federrath2012}.
In more extreme environments, the deviation from the standard SF relations can be even more severe.
Some high-redshift disc and starburst galaxies have been reported to have larger $\epsilon_\mathrm{SF}$ by a factor of at least 10 
\citep[see e.g.][]{Daddi2010,Genzel2010} while in the MW's Central Molecular Zone (CMZ),
a possible analog for high redshift star formation environments,
the efficiency appears to be a factor of 10-100 lower (see e.g. \citealt{Longmore2013}; however, see e.g. \citealt{Sharda2018} and 
\citealt{Federrath2016} for applications of (magneto)-turbulent SF models to high redshift and MW CMZ environments, respectively).
Additionally, in certain circumstances it is possible that while an average over long timescales yields some value of the efficiency,
it may in fact vary episodically on smaller scales, either due to regulation by turbulent pressure \citep[see e.g.][]{Kruijssen2014}
or by feedback regulation \citep[see e.g.][]{Grudic2018}. 
The upshot of both of these scenarios is that using a large
spatial scale and `long' timescale averaged value of $\epsilon_\mathrm{SF}$ may artificially 
smooth out star formation and, crucially for this work, SN rates.

The reason for the increase in SN feedback efficiency as a result of increasing
$\epsilon_\mathrm{SF}$ is twofold. Firstly, it leads to more clustered SNe
that are able to work together to drive outflows. Secondly, it avoids the
issue of building up high density gas by efficiently converting gas into stars
before the problem arises. Care, however, must be taken when using such high
values of the efficiency that this does not represent an unphysical removal of
gas. As the gas consumption time is then effectively the free-fall time, above $100\ \mathrm{cm^{-3}}$ this becomes comparable to the time before the
first SNe explode, meaning that most, if not all, of the local gas will have
been converted into stars, significantly dropping local density for subsequent
SN events. If the internal structure of star-forming regions is well
resolved this may not particularly problematic because the
hydrodynamics should correctly follow the fragmentation of the region
without recourse to `fudge factors'. However, if the region is unresolved,
using an efficiency of $100\%$ will quickly convert the entire mass of the
region into stars, which is likely unphysical. In other words, if we are confident
that we fully resolve all the relevant small scale processes and timescales (for example,
that our hydrodynamics will correctly capture effects such as turbulent support, or that our subgrid feedback
prescriptions are fully physical), then we can use a high star formation efficiency
coupled with some smaller scale restrictions for which gas can form stars (e.g. virial
parameter, Jeans unstable gas etc.)
and rely on these processes to correctly regulate the resulting SFRs. If not,
then the results are likely to be erroneous. For example, in a scenario
such as that described by \cite{Kruijssen2014}, if we fail to resolve the turbulent
pressure (and other relevant small scale details) that leads to episodic star formation, then we will entirely miss the
low efficiency section of the cycle. In our case, it is likely that
we sit somewhere in between these two cases. Our fiducial choice of  a fixed $\epsilon_\mathrm{SF}=1.5\%$
is possibly too conservative. On the other hand, it is not clear that
we capture the small scale structure and turbulence of the ISM sufficiently to
justify 100\%, probably leading to the unphysically rapid consumption of star forming
regions by gas `deletion' and subsequently overpowered SN feedback. It should be noted
that roughly this magnitude of $\epsilon_\mathrm{SF}$ will be required to regulate
SF, as using 10 times our fiducial value also failed to regulate star formation.

Furthermore, we have experimented with the adoption of a SF prescription that uses
a variable efficiency based on local turbulent gas properties (with a prescription similar to \citealt{Kimm2017}).
This scheme attempts to infer the likely (unresolved) turbulent Mach number, $\mathcal{M}$,
and virial parameter, $\alpha$, based on the resolved local velocity gradients. These are then used as inputs into 
the analytic star formation law of \cite{Padoan2011} \citep[see also this formalism explored in][]{Federrath2012}.
This derives a star formation efficiency per freefall time by calculating the fraction of gas above some critical density,
determined by considering the particular log-normal density distribution of gas expected for the given values of $\mathrm{M}$
and $\alpha$. We leave a detailed discussion to a
future work (Smith et al. 2019 in prep.), but find it worthwhile to report the tentative result that
in this specific case there is little impact on the evolution of Dwarf 1. This is largely because we find in our
simulations these models typically give $\epsilon_\mathrm{SF}\approx1\%-20\%$, which we have already demonstrated is not
high enough to sufficiently
enhance the strength of SN feedback such that it makes a difference to the evolution of our dwarfs. 
Nonetheless, it is clear that the efficiency of SN feedback is
strongly dependent on their spatial and temporal clustering. Since this is
explicitly tied to the manner in which star formation proceeds on local scales, it is
important to model this in a physical a manner as possible.

Another phenomenon which impacts SN clustering properties is the fraction 
of walkaway/runaway SN progenitors. Dynamical interactions
during the formation of a star cluster may eject progenitors \citep[see e.g.][]{Poveda1967,Fujii2011,Oh2015}
or alternatively runaways may be caused by the occurrence of a SNe in an OB binary system
(see e.g. \citealt{Blaauw1961,PortegiesZwart2000,Eldridge2011}, see also \citealt{Kim2017} for a subgrid implementation
of this mechanism).
If a progenitor is able to travel away from its
birth site the subsequent SN is more likely to occur in a low density medium which
maximises its efficiency. Conversely, a high fraction of runaways will tend to smooth
out the spatial clustering of the ensemble of SNe, potentially reducing the ability
of remnants to overlap and form superbubbles. Finally, if SNe occur outside of the
dense star forming clouds they may not be able to efficiently disperse star forming
gas. Speculating on the dominant impact of runaway SN progenitors is beyond the scope
of this present work (given that it is likely to be sensitive to the exact parameters 
adopted such as runaway fraction, velocity distribution etc.), but we note that they may play an important role in determining
overall SN feedback efficiency.

It is worth reemphasizing that regardless of the star formation criteria, there
is a large body of theoretical and observational work indicating that
other sources of stellar feedback must be operating prior to the first SN,
such as stellar winds, photoelectric heating and photoionization from young
stars. These processes may have a significant impact on local gas,
not only affecting its density and temperature structure, but also the level
of turbulent support. Given that we have demonstrated a tendency for dense gas
to build up and overwhelm SN feedback in our $z=4$ dwarfs (and that this
effect is physically realistic, rather than just being a symptom of numerical
overcooling), it may be the case that non-SN stellar feedback plays a more
important role in the evolution of low mass haloes than is commonly
assumed. This conclusion is consistent with the results found by the
\textsc{FIRE-1} project \citep{Hopkins2014a} in which the removal of other sources
of stellar feedback in dwarfs led to SN feedback having almost no impact on
stellar mass (though the effect appears to be less severe in \textsc{FIRE-2}
\citep{Hopkins2017a}). Finally, we note that the efficiency of first (and subsequent) SN
events may depend on the fraction of runaway SN and on alternative heating
processes such as those provided by relativistically accelerated particles in
the wake of SN explosions. 

\vspace{-4ex}
\section{Conclusion}
We have carried out very high resolution cosmological zoom-in simulations of
five dwarf galaxies up to $z = 4$ with virial masses between $\sim
2-6\times10^9\ \mathrm{M_\odot}$. Our simulations adopt the mechanical SN
feedback scheme introduced in \cite{Smith2018} and a spatially constant, but
time evolving UV background \citep{FG2009}. The SN feedback is constructed to
deliver the correct momentum to the surrounding ISM corresponding to the stage
of the SN remnant evolution. We found that this model leads to self-regulated star
formation rates, realistic galaxy kinematics and gas content thanks to the
occurrence of multiphase, mass-loaded outflows in isolated dwarf simulations
\citep{Smith2018}. The aim of the present work is to determine whether the
same model of SN feedback results in the realistic dwarfs properties
once the full cosmological formation is incorporated self-consistently. We
find that:  
\begin{itemize}
\item Without the inclusion of SN feedback, we produce dwarfs that have over
  3 orders of magnitude too much stellar mass relative to (extrapolated)
  abundance matching predictions. Their stellar and gas metallicities are in
  excess of solar abundances. The dwarfs undergo a catastrophic collapse to
  the resolution limit, resulting in extremely dense systems with strongly
  peaked circular velocity curves. Dark matter density in the centre of the
  halo is enhanced relative to a collisionless simulation by approximately an
  order of magnitude.

\item In general, while the inclusion of SN feedback induces more bursty SFR
  rates and affects dwarf morphologies, it has insufficient impact on the total
  stellar mass formed. In the majority of our systems, the build up of dense
  gas (often following a wet merger) renders the SNe too inefficient to expel
  gas from the galaxy and suppress star formation. We emphasise because
  our scheme injects the correct amount of momentum per SN, this
  effect is not an example of classical numerical overcooling but rather a
  physical suppression of SN efficiency. Most SNe explode in gas of density
  $10^4\ \mathrm{cm^{-3}}$ which limits the feedback momentum budget
  available. This suggests that some other mechanism(s) must be invoked
  (e.g. other sources of stellar feedback) that can prevent gas from
  collapsing to such high densities and/or clear it prior to SNe occurring. Inclusion of
  runaway SN may help alleviate this issue as well.

\item We however find one exception to this scenario where we are able to
  produce a realistic dwarf relative to the extrapolations of abundance matching
  and various metrics of local analogs. Our dwarf 4 forms by a major merger
  relatively late in its history at $z \approx 5.5$. It therefore spends most of
  its evolution as two lower mass systems in which the SNe are able to expel
  gas and halt star formation before catastrophic collapse sets in. Their late
  major merger is therefore mostly dry and does not trigger more than a brief
  burst of star formation which is quickly suppressed by feedback. We note
  that while SNe feedback is clearly efficient here, enriching the CGM to a few
  $10^{-2}\ \mathrm{Z_\odot}$ with mass-loaded winds, no prominent dark matter
  core forms.

\item We have carried out a variety of other simulations to test the
  applicability of our conclusions. We find that our results are not
  significantly impacted by increasing resolution, changing details of the
  (spatially uniform) UV background or removing the pressure floor. Our
  results are also relatively insensitive to increasing the star formation density
  threshold by an order of magnitude. Arbitrarily increasing the star
  formation efficiency parameter by an order of magnitude to 15\% leads to
  more bursty behaviour and reduced star formation, but still overshoots
  abundance matching relations by 2 orders of magnitude. Only by taking an
  extreme choice of parameters, using a star formation efficiency of $100\%$, 
  are we able to get close to the relation. 
\end{itemize}  

We have demonstrated that realistically modelled SN feedback is easily
overwhelmed early on in the cosmological assembly of dwarfs by the
build up of gas, despite the relatively shallow potential well. While this can
potentially be dealt with by adopting a star formation prescription that leads
to extremely concentrated SN feedback, it seems that some combination of other sources of
stellar feedback and/or currently unresolved turbulent support may be required to modulate ISM densities prior to the first
SNe exploding in order to preserve their efficiency.

\vspace{-4ex}
\section*{Acknowledgements}
We are grateful to Cathie Clarke, Adrianne Slyz, Christoph Federrath and the anonymous referee for helpful
comments. MCS and DS acknowledge support by the Science and Technology Facilities Council (STFC) and
the ERC Starting Grant 638707 ``Black holes and their host galaxies:
co-evolution across cosmic time''. This work
was performed on the following: the DiRAC Data Analytic system at the University of Cambridge,
operated by the University of Cambridge High Performance Computing Service on behalf of the STFC
DiRAC HPC Facility (www.dirac.ac.uk). This equipment was funded by BIS National E-infrastructure capital grant (ST/K001590/1), STFC capital grants ST/H008861/1 and ST/H00887X/1, and STFC DiRAC Operations grant ST/K00333X/1; the Cambridge Service for Data Driven Discovery (CSD3), part of which is operated by the University of Cambridge Research Computing on behalf of the STFC DiRAC HPC Facility. The DiRAC component of CSD3 was funded by BEIS capital funding via STFC capital grants ST/P002307/1 and ST/R002452/1 and STFC operations grant ST/R00689X/1; the DiRAC@Durham facility managed by the Institute for Computational Cosmology on behalf of the STFC DiRAC HPC Facility. The equipment was funded by BEIS capital funding via STFC capital grants ST/P002293/1 and ST/R002371/1, Durham University and STFC operations grant ST/R000832/1. DiRAC is part of the National e-Infrastructure.

\vspace{-4ex}
\bibliographystyle{mn2e} 
\bibliography{references}

\end{document}